\documentclass[prx,twocolumn,english,floatfix,longbibliography]{revtex4-1} 

\usepackage{graphicx}
\usepackage{dcolumn}
\usepackage{bm}
\usepackage{amsmath,amssymb}
\usepackage{amsthm}
\usepackage{mathrsfs}
\usepackage{indentfirst}
\usepackage{float}
\usepackage{braket}
\usepackage[utf8]{inputenc}
\usepackage{bm}
\usepackage{fancyhdr}
\usepackage{graphicx}
\usepackage{tikz}
\usepackage{physics}
\usepackage{natbib}
\usetikzlibrary{arrows, shapes.gates.logic.US, calc}
\usetikzlibrary{angles,quotes}
\tikzstyle{branch}=[fill, shape=circle, minimum size=3pt, inner sep=0pt]
\usepackage{blochsphere}
\usepackage{mathtools}
\usepackage[normalem]{ulem}

\begin{document}

\preprint{APS/123-QED}

\title{
Fundamental limits of superconducting quantum computers}

\author{Michele Vischi$^{1,2}$, Luca Ferialdi$^{1,2}$, Andrea Trombettoni$^{1,2,3}$, Angelo Bassi$^{1,2,4}$}
\affiliation{$^{1}$
Department of Physics, University of Trieste, Strada Costiera 11, 34151 Trieste, Italy}
\affiliation{$^{2}$Istituto Nazionale di Fisica Nucleare, Trieste Section, Via Valerio 2, 34127 Trieste, Italy}
\affiliation{$^{3}$CNR-IOM DEMOCRITOS Simulation Center, Via Bonomea 265, I-34136 Trieste, Italy}
\affiliation{$^{4}$National Institute of Optics - CNR - Research Unit of Trieste, Strada Statale 14 - 34149 Trieste}




\date{\today}

\begin{abstract}
The Continuous Spontaneous Localization (CSL) model is an alternative formulation of quantum mechanics which introduces a noise coupled non linearly to the wave function to account for its collapse. We consider CSL effects on quantum computers made of superconducting transmon qubits. As a direct effect CSL reduces quantum superpositions of the computational basis states of the qubits: we show the reduction rate to be negligibly small. However, an indirect effect of CSL, dissipation induced by the noise, also leads transmon qubits to decohere, by generating additional quasiparticles. Since the decoherence rate of transmon qubits depends on the quasiparticle density, by computing their generation rate induced by CSL, we can estimate the corresponding quasiparticle density and thus the limit set by CSL on the performances of transmon quantum computers. We show that CSL could spoil the quantum computation of practical algorithms on large devices. We further explore the possibility of testing CSL effects on superconducting devices.
\end{abstract}

\maketitle

\section{\label{sec:level1}Introduction}
Quantum computers bring with themselves the promise to allow for a significant speedup in the resolution of many relevant complex problems with respect to current classical computers \cite{shor1999polynomial,brassard2002quantum,montanaro2015quantum}. When large enough quantum computers will be available, they are expected to impact many fields such as cybersecurity \cite{shor1999polynomial}, drug synthesis \cite{cao2018potential}, simulation of quantum systems \cite{childs2018toward}, to name a few. 
Such computers will need a large processor comprising many qubits, the fundamental units of quantum computation (the analog of bits in classical computers). 
Different physical realizations of qubits are currently under development, e.g. superconducting qubits \cite{kjaergaard2020superconducting}, trapped ions \cite{bruzewicz2019trapped}, photonic chips \cite{bombin2021interleaving} and spin qubits \cite{recher2010quantum}. Superconducting qubits, which are electrical circuits made of superconducting materials, are among the most promising for scaling up quantum processors.

Quantum chips made of so called transmon superconducting qubits \cite{koch2007charge} are currently fabricated by the major companies investing in quantum computing, such as Google \cite{arute2019quantum}, IBM \cite{gambetta2020ibm} and Rigetti \cite{sete2016functional}. State of the art transmon devices contain $\sim 10^2$ qubits, and more than once were used to reach quantum supremacy \cite{arute2019quantum,wu2021strong}, i.e. the resolution of a problem faster than what possible with any classical computer. Although this is an outstanding result, the problems solved so far do not have any practical application.
Many technical hurdles must be overcome for such quantum machines to be practical. The main challenge to address is the fragility of quantum states that are stored in quantum computers: even a single qubit suffers decoherence, i.e. the undesired loss of its quantum properties over time.

Several sources of noise lead to decoherence. We can divide them into two categories: environmental noise sources, whose effects can be eventually mitigated with technological development, and possible fundamental noise sources, that are instead unavoidable. Most of the research focuses on how to protect qubits from environmental noises \cite{corcoles2011protecting,wang2015surface,klimov2018fluctuations,catelani2011relaxation,glazman2021bogoliubov,martinis2009energy}, and less attention has been paid to identifying fundamental ones \cite{lloyd2000ultimate,gambini2005fundamental}. While technology develops, we may reach a level of control and accuracy at which fundamental noises can not be overlooked anymore. Among these, of particular relevance are those related to the spontaneous collapse of the quantum states, as predicted by the continuous spontaneous localization (CSL) model \cite{bassi2003dynamical}.

The CSL model (more generally, collapse models \cite{bassi2003dynamical}) is an alternative formulation of quantum mechanics developed in order to solve the tension between the quantum superposition principle and the wave packet reduction postulate.
At the core of the model there is a classical noise, suitably coupled to the wave function of particles accounting for its collapse, which now becomes part of the dynamics, not a separate postulate.
The CSL state vector reduction of quantum states becomes more effective as the difference in mass density of the states in superposition increases \cite{ferialdi2020continuous}. In this way the model is consistent with quantum mechanics in the microscopic regime, where the standard theory gives extremely accurate predictions, at the same time justifying why macroscopic objects are always localized in space.

Collapse models set an intrinsic limit to the stability of quantum systems over time. As a consequence, they are expected to set a limit to the scalability of quantum computers, which is the subject of this work where we will consider quantum computers based on superconducting technology.
As we will see, the CSL state vector reduction does not limit significantly the performance of transmon qubits. However an indirect effect of the collapse, dissipation, is more relevant, given the extreme sensibility of superconducting devices to perturbations. We will compute this effect and will show that, when neglecting other sources of noise is (or will be) possible, dissipation can be detected at current experimental temperatures. Moreover we will see that CSL could spoil the quantum computation of complex practical algorithms on large devices. 


The paper is organized as follows. In section \ref{BCS} we review for convenience the key elements of the Bardeen-Cooper-Schrieffer (BCS) theory of superconductors needed for the subsequent analysis, highlighting the role played by the excited states of a superconductor, the quasiparticles. 
In section \ref{csl} we discuss how to treat CSL noise in the BCS formalism. The results provide the basis for the study of the effect of CSL terms in superconducting devices and computers. We will then outline the key elements of the CSL model that are needed to compute CSL effects on transmon qubits.
In section \ref{reduction} we focus on transmon qubits and we compute the reduction rate of CSL, i.e. the rate at which CSL localizes a superposition of transmon qubits. We estimate that this effect is at date negligibly small for practical purposes.
We then characterize the effects of the CSL dissipation on transmon devices. In section \ref{trans} we show how the CSL noise perturbs the superconducting materials in transmon qubits, by generating additional quasiparticles. These excited states accumulate over time in the devices' volume leading to a steady state quasiparticle density larger than the thermal one. This excess limits the coherence time of transmon qubits. We perform the computation of the CSL quasiparticle density in section \ref{phonon}. In section \ref{testingCSL}, we explore the possibility of testing collapse models with superconducting devices, showing the experimental conditions to be met in order to detect the CSL excess of quasiparticles.
In section \ref{fundmentallimits} we estimate the fundamental limit due to CSL on the performances of transmon quantum computers given the quasiparticle density computed in section \ref{phonon}. 
In section \ref{conclusion} we conclude with final remarks and outlook.


\section{BCS theory}\label{BCS}
Transmon qubits exploit the peculiar properties of superconducting materials. This section is devoted to introducing the key concepts and quantities of the microscopic theory of superconductors that will be used in the following.

Conventional superconductors are described by the BCS theory. Below a critical temperature $T_c$, it is energetically convenient for electrons in some metals to bind in pairs, called Cooper pairs. The attraction is mediated by phonons, and the two electrons in a Cooper pair have opposite momenta and spins. The total spin is zero, which allows different pairs to behave coherently similarly to (but not exactly as) boson condensates. 

The BCS Hamiltonian of the system of electrons is given by \cite{schrieffer2018theory,grosso2013solid}:
\begin{equation}\label{bcsham}
    \hat{H}_{BCS} = \sum_{\mathbf{k}\sigma} \xi_{\mathbf{k}} \hat{c}_{\mathbf{k}\sigma}^{\dagger}\hat{c}_{\mathbf{k}\sigma} + \sum_{\mathbf{k}\mathbf{k}'} U_{\mathbf{k}\mathbf{k}'}\hat{c}_{\mathbf{k}\uparrow}^{\dagger}\hat{c}_{-\mathbf{k}\downarrow}^{\dagger}\hat{c}_{-\mathbf{k}'\downarrow}\hat{c}_{\mathbf{k}'\uparrow}
\end{equation}
where $\xi_{\mathbf{k}} = \hbar^2 k^2/2m -\epsilon_F$ is the energy measured with respect to the Fermi energy $\epsilon_F$, and $U_{\mathbf{k}\mathbf{k}'}$ are the matrix elements of the interaction potential. The first term of the Hamiltonian is the kinetic energy while the second potential term couples pairs of different momenta $k$ and $k'$.
The ground state of a superconductor is given by the BCS ground state:
\begin{equation}\label{gsbcs}
    \ket{\psi_{S}} = \prod_{\mathbf{k}} (u_{\mathbf{k}}+e^{i\phi}v_{\mathbf{k}}\hat{c}_{\mathbf{k}\uparrow}^{\dagger}\hat{c}_{-\mathbf{k}\downarrow}^{\dagger})\ket{0}\, ,
\end{equation}
where $\ket{0}$ is the vacuum state of the electrons and the operators $\hat{c}^{\dagger}_{\mathbf{k}\sigma}$ ($\hat{c}_{\mathbf{k}\sigma}$) create (destroy) an electron of momentum $k$ and spin $\sigma$. The real coefficients $u_{\mathbf{k}}$ and $v_{\mathbf{k}}$ satisfy the normalization conditions $u_{\mathbf{k}}^2+v_{\mathbf{k}}^2 = 1$. $v_{\mathbf{k}}^2$ ($u_{\mathbf{k}}^2$) gives the probability that the Cooper pair of momentum $k$ is occupied (unoccupied).

Since finding the excited states of a superconductor in terms of electron operators is not easy, it is convenient to perform a Bogoliubov transformation. Such a transformation diagonalizes the Hamiltonian (\ref{bcsham}) by introducing new canonical fermionic operators $\hat{\gamma}_{\mathbf{k}\sigma}$ (see appendix \ref{firstapp}):
\begin{equation}
    \hat{H}_B = \sum_{\mathbf{k}\sigma} E_{\mathbf{k}} \hat{\gamma}^{\dagger}_{\mathbf{k}\sigma}\hat{\gamma}_{\mathbf{k}\sigma}\, ,
\end{equation}
where $E_{\mathbf{k}}$ is the energy of a an excited state associated to momentum $k$:
\begin{equation}\label{qpenergies}
    E_{\mathbf{k}} = \sqrt{\xi_{\mathbf{k}}^2+\Delta_{\mathbf{k}}^2}\, ,
\end{equation}
and $\Delta_{\mathbf{k}}$ are the so called superconducting gap parameters. 
The operators $\hat{\gamma}_{\mathbf{k}\sigma}$ are such that the BCS ground state is their vacuum state, i.e. $\hat{\gamma}_{\mathbf{k}\sigma}\ket{\Psi_{S}} = 0$. Acting with $\hat{\gamma}_{\mathbf{k}\sigma}^{\dagger}$ on the BCS ground state gives an excited state, called quasiparticle. 
The excited states in terms of $\hat{c}^{\dagger}_{\mathbf{k}\sigma}$ read:
\begin{align}
        & \ket{\psi_1}=\hat{\gamma}_{\mathbf{k}\uparrow}^{\dagger}\ket{\psi_S} = \hat{c}_{\mathbf{k}\uparrow}^{\dagger}\prod_{\mathbf{l}\neq \mathbf{k}}(u_{\mathbf{l}}+ v_{\mathbf{l}} \hat{c}_{\mathbf{l}\uparrow}^{\dagger}\hat{c}_{-\mathbf{l}\downarrow}^{\dagger})\ket{0} \, \\
        &       \ket{\psi_2}=\hat{\gamma}_{-\mathbf{k}\downarrow}^{\dagger}\ket{\psi_S} = \hat{c}_{-\mathbf{k}\downarrow}^{\dagger}\prod_{\mathbf{l}\neq \mathbf{k}}(u_{\mathbf{l}} + v_{\mathbf{l}} c_{\mathbf{l}\uparrow}^{\dagger}c_{-\mathbf{l}\downarrow}^{\dagger})\ket{0}\, ,
\end{align}
which means that for given momentum $k$ there is an electron with probability $1$ and the other state of the pair is empty.
Quasiparticles can be then interpreted as fermions created by $\hat{\gamma}^{\dagger}_{\mathbf{k}\sigma}$ which are in one-to-one correspondence with the $\hat{c}^{\dagger}_{\mathbf{k}\sigma}$. 

The coefficients $u_{\mathbf{k}}$ and $v_{\mathbf{k}}$ are equal to:
\begin{align}\label{ukvk}
    & u_{\mathbf{k}}^2 =\frac{1}{2}\bigg(1+\frac{\xi_{\mathbf{k}}}{E_{\mathbf{k}}}\bigg) &v_{\mathbf{k}}^2=\frac{1}{2}\bigg(1-\frac{\xi_{\mathbf{k}}}{E_{\mathbf{k}}}\bigg) \, .
\end{align}
One can see that, as $k$ varies from well below the Fermi surface to well above it, $v_{\mathbf{k}}^2$ goes from $1$ to $0$ (and analogously $u_{\mathbf{k}}^2$ goes from $0$ to $1$), i.e. Cooper pairs of momentum $k$ well below the Fermi surface are occupied in the ground state with probability $1$. As the momentum of the Cooper pairs increases above the Fermi surface, the occupation probability decreases to $0$. 

In general, the superconducting gap parameters $\Delta_{\mathbf{k}}$ have different values for different $k$.
As well known, BCS simplifies this (see appendix \ref{firstapp}) by setting $\Delta_{\mathbf{k}} = \Delta$ for values of $k$ such that $|\xi_{\mathbf{k}}|<\hbar\omega_D$ (with $\omega_D$ the Debye frequency of phonons), and $\Delta_{\mathbf{k}} = 0$ otherwise. 
$\Delta$ is 
the superconducting gap, and its meaning becomes clearer by looking at the quasiparticle energies $E_{\mathbf{k}}$ as function of $\xi_{\mathbf{k}}$ in Eq.(\ref{qpenergies}). They have a minimum at $\xi_{\mathbf{k}}= 0$ and the value of the excitation energy at this point is $\Delta$. This implies that to have an excited states at least an amount of energy $\Delta$ is required.

The superconducting gap is in general a function of temperature, but at sufficiently low temperatures (we will make this assumption hereafter) it is 
$\Delta(T) \approx \Delta(0) = 1.76k_B T_c$, where $T_c$ is the critical temperature and $k_B$ is the Boltzmann constant.

In the realistic case of a small, but non vanishing $T$, some quasiparticle will be thermally excited, and they will be distributed according to some occupation function $f(E,T)$. For a superconductor at thermal equilibrium the quasiparticles have a Fermi-Dirac occupation function:
\begin{equation}\label{fermidirac}
    f_{FD}(E,T) = \frac{1}{e^{E/k_BT}+1}\, .
\end{equation}
The normalized quasiparticle density $x_{qp}$, i.e. the ratio between the number of quasiparticles and the number of Cooper pairs inside a superconductor, gives an estimate of how many quasiparticles there are in a superconducting device. Its expression is given by:
\begin{equation}\label{qpdensity}
    x_{qp} = \int_{\Delta}^{\infty}f(E)\rho(E)dE\, ,
\end{equation}
where $\rho(E)$ is the normalized superconducting density of states:
\begin{equation}\label{scdensity}
   \rho(E) = \frac{E}{\sqrt{E^2-\Delta^2}} \, .
\end{equation}

Making the assumptions that $f(E)$ is a Fermi-Dirac distribution and that $x_{qp}$ is small, so that most of the states are close to the gap, one can find the following expression for $x_{qp}$ \cite{catelani2011quasiparticle}:
\begin{equation}\label{qpdensityexpr}
    x_{qp}=\sqrt{2\pi k_BT/\Delta}e^{-\Delta/k_BT}\, .
\end{equation}

Inserting in this equation temperatures close to the experimental regime of $20$ mK, and the parameters of aluminum (a typical superconductor used in transmon qubits), one can see that the quasiparticle density should be exponentially suppressed: $x_{qp}\sim 10^{-52}$ \cite{de2014evidence}. For the purposes of this paper, it is very important to stress that however experiments on superconducting qubits and superconducting resonators show higher density values, $x^{\text \tiny{exp}}_{qp}\sim 10^{-9}-10^{-6}$ \cite{serniak2019direct}. Thermal equilibrium seems not able to explain such an excess of quasiparticles and for this reason the latter are usually called non-equilibrium quasiparticles. Appendix \ref{firstapp} contains further technical details about the BCS theory.

\section{CSL model in the BCS framework}\label{csl}
The CSL model is an alternative formulation of quantum mechanics devised to solve the problem of the quantum-to-classical transition in quantum theory.
CSL unifies the Schrödinger evolution, which is linear and deterministic, with the non-linear and stochastic dynamics giving wave packet reduction. This is done by modifying the Schrödinger equation, adding stochastic and non-linear terms that implement the collapse of the wave function. This dynamical modification is consistent with quantum mechanics in the microscopic regime, where the standard theory gives extremely accurate predictions, at the same time justifying why macroscopic objects are always localized in space. The strength and spatial extension of the collapse (which is white in time) are dictated respectively by two parameters of the theory, $\lambda$ and $r_c$. Theoretical arguments suggest that $r_c \sim 10^{-7}$ m \cite{bassi2003dynamical,carlesso2019collapse} and $\lambda$ in the range $\sim 10^{-8}-10^{-16}$s$^{-1}$\cite{bassi2003dynamical,adler2007lower}. From now on, to fix a possible value we set $\lambda = 10^{-10}$s$^{-1}$. 

The full non-linear and stochastic CSL dynamics is not easy to work with. Since we are interested in prediction of observable effects, one can equivalently use the following simplified linear dynamics (see appendix \ref{secondapp} for further details):
\begin{equation}\label{evcslsimpl}
     i\hbar\frac{d\ket{\psi(t)}}{dt} = (\hat{H} +\hat{H}_\text{\tiny CSL})\ket{\psi(t)}\, ,
\end{equation}
where $\hat{H}$ is the Hamiltonian of the system ($\hat{H}_{BCS}$ in the present case) and $\hat{H}_\text{\tiny CSL}$ is the CSL contribution. By performing a Fourier transform to momentum space in a normalization box of volume $V$ to avoid potential divergences (see appendix \ref{secondapp} for the details), $\hat{H}_\text{\tiny CSL}$ takes the following expression:
\begin{equation}\label{cslhamc}
\begin{split}
   &\hat{H}_\text{\tiny CSL} = -\frac{m\hbar\sqrt{\lambda}}{m_0 V}\sum_{\mathbf{k_1}\mathbf{k_2},s}\widetilde{W}_{\mathbf{k_1}-\mathbf{k_2}}(t)\, \widetilde{G}_{\mathbf{k_1}-\mathbf{k_2}} \,\hat{c}^{\dagger}_{\mathbf{k}_1s}\hat{c}_{\mathbf{k}_2s}\,,
   \end{split}
\end{equation}
where $m_0$ is the nucleon mass and $m$ is the mass of the particle considered (in our case, the electrons).
The stochastic processes $\widetilde{W}_{\mathbf{k}}(t)$ have expectation value and two-point correlator given by:
\begin{align}
    & \mathbb{E}[\widetilde{W}_{\mathbf{k_1}}(t)]=0 \\ &\label{correlatortp}\mathbb{E}[\widetilde{W}_{\mathbf{k_1}}(t)\widetilde{W}_{\mathbf{k_2}}(s)]= V \delta_{(\mathbf{k_1} + \mathbf{k_2})}\delta(t-s)\, ,
\end{align}
and they are weighted by the Gaussian function:
\begin{equation}
   \widetilde{G}_{\mathbf{k}} = (4\pi r_c^2)^{3/4} e^{-\frac{r_c^2 k^2}{2}}\, .
\end{equation} 

Equation (\ref{cslhamc}) shows that the CSL noise scatters electrons, effectively acting as a kick which adds energy to the system. 
The goal of the present analysis is to quantify the main effect of those kicks on the BCS ground state of a superconductor, which, as we will see, amounts to the breaking of Cooper pairs into quasiparticles.
More precisely, we will compute the transition probability from a superconductor initially in its ground state to states containing quasiparticles. We will use perturbation theory, thus considering the CSL Hamiltonian (\ref{cslhamc}) as an interacting term with the diagonal Bogoliubov Hamiltonian $\hat{H}_B$ as the free term.

In this setup, the transition probability from the initial state $\ket{i}$ to a final state $\ket{f}$ is given by:
\begin{equation}\label{transprobboh}
P_{fi} =\mathbb{E}[|T_{fi}|^2]= \mathbb{E}[|\bra{f}\hat{U}_I(t,t_i)\ket{i}|^2] \,, 
\end{equation}
where $T_{fi}$ is the transition amplitude from $\ket{i}$ to a state $\ket{f}$ and $\hat{U}_I(t,t_i)$ is the time evolution operator in the interaction picture:
\begin{equation}
    \hat{U}_I(t,t_i) = e^{\frac{i}{\hbar}\hat{H}_B t}\hat{U}(t,t_i)e^{-\frac{i}{\hbar}\hat{H}_B t}\, .
\end{equation}
The operator $\hat{U}_I(t,t_i)$ can be expanded in the Dyson series, which at first order reads:
\begin{equation}\label{dyson}
    \hat{U}(t,t_i) = 1 + \int_{t_i}^{t} ds \,\hat{H}^{(I)}_\text{\tiny CSL}(s) \, .
\end{equation}
where $\hat{H}^{(I)}_\text{\tiny CSL}(t) = e^{\frac{i}{\hbar}\hat{H}_B t}\hat{H}_\text{\tiny CSL}e^{-\frac{i}{\hbar}\hat{H}_B t}$ is the Hamiltonian (\ref{cslhamc}) in the interaction picture. 

The diagonal form of the Bogoliubov Hamiltonian in terms of $\hat{\gamma}_{\mathbf{k}\sigma}$ and the properties of the BCS ground state for these operators make it easier to work with creation and annihilation operators for quasiparticles $\hat{\gamma}^{\dagger}_{\mathbf{k}\sigma}$ and $\hat{\gamma}_{\mathbf{k}\sigma}$.
Using the inverse Bogoliubov transformation [see Eq.(\ref{boginverse}) in appendix \ref{firstapp}], we can transform the CSL Hamiltonian in terms of the operators $\hat{\gamma}_{\mathbf{k}s}$ and then find its expression in the interaction picture, reading:
\begin{equation}\label{cslhamgamintmain}
\begin{split}
   &\hat{H}^{(I)}_\text{\tiny CSL} = -\frac{\hbar\sqrt{\lambda}m}{m_0 V}\sum_{\mathbf{k_1}\mathbf{k_2}}\widetilde{G}_{\mathbf{k_1}-\mathbf{k_2}}\times\\
   &\bigg[\widetilde{W}_{\mathbf{k_1}-\mathbf{k_2}}(t)L(k_1,k_2)e^{\frac{i}{\hbar}(E_{\mathbf{k}_1}-E_{\mathbf{k}_2})t}\hat{\gamma}^{\dagger}_{\mathbf{k}_1\uparrow}\hat{\gamma}_{\mathbf{k}_2\uparrow}\\
   &+\widetilde{W}_{\mathbf{k_2}-\mathbf{k_1}}(t)L(k_1,k_2)e^{\frac{i}{\hbar}(E_{\mathbf{k}_2}-E_{\mathbf{k}_1})t}\hat{\gamma}^{\dagger}_{-\mathbf{k}_2\downarrow}\hat{\gamma}_{-\mathbf{k}_1\downarrow}\\
   &+\widetilde{W}_{\mathbf{k_2}-\mathbf{k_1}}(t)M(k_1,k_2)e^{-i\phi}e^{-\frac{i}{\hbar}(E_{\mathbf{k}_1}+E_{\mathbf{k}_2})t}\hat{\gamma}_{\mathbf{k}_2\uparrow}\hat{\gamma}_{-\mathbf{k}_1\downarrow}\\
   &+\widetilde{W}_{\mathbf{k_1}-\mathbf{k_2}}(t)M(k_1,k_2)e^{i\phi}e^{\frac{i}{\hbar}(E_{\mathbf{k}_1}+E_{\mathbf{k}_2})t}\hat{\gamma}^{\dagger}_{\mathbf{k}_1\uparrow}\hat{\gamma}^{\dagger}_{-\mathbf{k}_2\downarrow}\bigg]  \, .
\end{split}
\end{equation}
The first two terms in the square brackets are associated to quasiparticle scattering: a quasiparticle of momentum $k_1$ is annihilated and another one of momentum $k_2$ is created. The third term is associated to quasiparticle recombination: two quasiparticles of different momenta are annihilated. The fourth term is the inverse process, called quasiparticle generation: two quasiparticle of different momenta are created. The functions $M$ and $L$, usually called coherence factors, are given by:
\begin{align}\label{coherencefact}
    & L(k_1,k_2) = (u_{\mathbf{k}_1}u_{\mathbf{k}_2}-v_{\mathbf{k}_1}v_{\mathbf{k}_2}) \\
    & M(k_1,k_2) = (u_{\mathbf{k}_1}v_{\mathbf{k}_2}+v_{\mathbf{k}_1}u_{\mathbf{k}_2}) \, .
\end{align}
Substituting Eqs. (\ref{ukvk}) we have that:
\begin{align}
    & L^2(E_1,E_2) = \frac{1}{2}\bigg(1-\frac{\Delta^2-\xi_{1}\xi_{2}}{E_{1}E_{2}}\bigg) \\
    & M^2(E_1,E_2) = \frac{1}{2}\bigg(1+\frac{\Delta^2-\xi_{1}\xi_{2}}{E_{1}E_{2}}\bigg)\, ,
\end{align}
where according to Eq.(\ref{qpenergies}), $\xi_{\mathbf{k}} = \sqrt{E_{\mathbf{k}}^2-\Delta_{\mathbf{k}}^2}$.

\section{CSL reduction on transmon qubits}\label{reduction}
As anticipated, the direct effect of CSL on superconducting quantum computers is to destroy superposition states. We now give an estimate of the rate at which CSL reduces superpositions of the computational basis states (labeled by $\ket{0}$ and $\ket{1}$) in a transmon qubit.
To estimate this rate, we first need a description of the physical form of the computational basis states of transmon qubits, keeping in mind that the CSL collapse mechanism is sensitive only to space superpositions of different masses.

A transmon qubit can be described as an electrical circuit in which a superconducting island (usually made of aluminum) of volume $V\sim 10^2\mu$m$^3$ is linked to a superconducting reservoir (of approximately the same size) through an insulating barrier of width $d \sim 1-10^2$ nm \cite{wang2014measurement}, forming a Josephson junction. By applying a gate voltage, Cooper pairs can tunnel through the Josephson junction from the reservoir into the island. The dimensions and the components of the transmon circuit are such that the number of excess Cooper pairs in the island becomes a quantum number, and the tunneling of a single Cooper pair can be controlled by manipulating the gate voltage.

The computational basis states of the transmon qubit are characterized by the number of excess Cooper pairs that have tunneled from the reservoir into the island.
The transmon qubit works in a regime for which $\ket{0}$ and $\ket{1}$ are not exactly eigenstates of the number of extra Cooper pairs operator, but rather both $\ket{0}$ and $\ket{1}$ are a superposition of states with a different number of extra Cooper pairs on the island. Nevertheless, for our estimate we can assume that, for a typical device, the difference in the number of pairs of the two computational basis states is of the order of $4$ \cite{koch2007charge}.
For our purposes, we can then effectively identify $\ket{0}$ as the state with $4$ Cooper pairs on one side of the Josephson junction, the reservoir, and $\ket{1}$ as the state with $4$ Cooper pairs on the other side of the junction, the island. 
We can think of these four Cooper pairs as two groups of four electrons. The two groups are separated by the BCS coherence length $\xi_c$ (a measure of the average distance between the two electrons in a Cooper pair) which for aluminum, the superconducting material used inside transmon circuits, is $\sim 10^{-6}$ m \cite{grosso2013solid}. 
Given these assumptions, we can compute the reduction rate with the formula \cite{adler2007lower}:
\begin{equation}\label{redcsladler}
    \Gamma_{R} =\lambda n^2 N\bigg(\frac{m_e}{m_0}\bigg)^2
\end{equation}
for $n$ particles within a radius smaller than the correlation length $r_c$, $N$ groups of particles separated by more than the correlation length $r_c$, and with $m_e$ the mass of electrons. 
In our effective model, if we substitute $n = 4$ and $N=2$ into Eq.(\ref{redcsladler}), considering an optimal configuration with four electrons packed within a distance smaller than $r_c$ in each of the $N=2$ groups of electrons separated by a distance $\xi_c>r_c$, we find the reduction rate $\Gamma= 32 \lambda (m_e/m_0)^2 \approx 10^{-16}$s$^{-1}$. We chose the value for $n$ that gives the highest value for $\Gamma_{R}$, to estimate the strongest theoretical CSL effect.
This shows that CSL reduction is negligibly weak, as the lifetime of a single qubit would be of the orders of billions of years. Even a large quantum computer composed by millions or even billions of such qubits, would be safe against the localization of superpositions dictated by collapse models.

However, besides collapsing superposition states, the CSL noise induces also dissipation as any other environmental noise: it perturbs the transmon qubits materials, also leading to decoherence effects. In fact, dissipation generates excited quasiparticle states that accumulate over time and their presence destroys quantum superpositions of transmon qubits.

\section{\label{sec:level2}Transition probability due to CSL dissipation}\label{trans}
CSL is ineffective in directly suppressing superposition states of superconducting qubits, mainly because too few electrons are involved in the superposition, which moreover have a very small mass. However, CSL impacts superpositions of superconducting qubits also indirectly.
As discussed in section \ref{csl}, the CSL noise couples to the Cooper pairs inside a superconductor, generating its excited states, the quasiparticles. Since the rate of decoherence of transmon qubits is proportional to the density of quasiparticles \cite{catelani2011quasiparticle}, these limit their performance. We now quantify the quasiparticle density due to CSL, to infer a limit on the coherence time of transmon qubits. To do so we compute the CSL transition probability to quasiparticle states, that will be used in the next section to obtain the CSL generation rate of quasiparticles and the evolution of the occupation function of quasiparticles $f(E)$.

We start by computing the transition probability in Eq.(\ref{transprobboh}) with $\ket{i} = \ket{\psi_S}$, using the time evolution operator expanded at first order as in Eq.(\ref{dyson}), with the CSL Hamiltonian in Eq.(\ref{cslhamgamintmain}), expressed in terms of quasiparticle operators. By recalling that $\ket{\psi_S}$ is the vacuum state for the operators $\hat{\gamma}_{\mathbf{k}s}$, only the fourth term of the Hamiltonian in Eq.(\ref{cslhamgamintmain}) gives a non vanishing contribution when acting on $\ket{\psi_S}$. This produces a transition to a final state different from $\ket{\psi_S}$, which contains quasiparticles. From these considerations, the main effect of the CSL noise on superconductors at first order in perturbation theory is the generation of quasiparticles. In particular, the only final states which give non zero contribution to the expectation value in (\ref{transprobboh}) are those of the form:
\begin{equation}\label{pq}
    \ket{f} = \hat{\gamma}^{\dagger}_{\mathbf{q}\uparrow}\hat{\gamma}^{\dagger}_{-\mathbf{p}\downarrow}\ket{\psi_S} \, ,
\end{equation}
where $q$ and $p$ are fixed.
Choosing $\ket{f}$ as in Eq.(\ref{pq}), the zero order term of the Dyson series gives a zero contribution when inserted into Eq.(\ref{transprobboh}).
The first order contribution of the Dyson series is:
\begin{equation}
    T^{(1)}_{qp} = -\frac{i}{\hbar}\int_{t_i}^{t} dt_1 \bra{\psi_S}\hat{\gamma}_{-\mathbf{p}\downarrow}\hat{\gamma}_{\mathbf{q}\uparrow}\hat{H}^{(I)}_\text{\tiny CSL}(t_1)\ket{\psi_S}\, ,
\end{equation}
and by substituting Eq.(\ref{cslhamgamintmain}) for the interacting Hamiltonian we find:
\begin{equation}
\begin{split}
    T^{(1)}_{qp} = &\frac{i\sqrt{\lambda}m}{m_0 V} \int_{t_i}^{t} dt_1 \, \sum_{\mathbf{k_1}\mathbf{k_2}}\widetilde{G}_{\mathbf{k_1}-\mathbf{k_2}}\widetilde{W}_{\mathbf{k_1}-\mathbf{k_2}}(t_1)M(k_1,k_2)\times\\
    &e^{i\phi}e^{\frac{i}{\hbar}(E_{\mathbf{k}_1}+E_{\mathbf{k}_2})t_1}\bra{\psi_S}\hat{\gamma}_{-p\downarrow}\hat{\gamma}_{q\uparrow}\hat{\gamma}^{\dagger}_{\mathbf{k}_1\uparrow}\hat{\gamma}^{\dagger}_{-\mathbf{k}_2\downarrow}\ket{\psi_S} \, .
\end{split}    
\end{equation}
The expectation value in the last line is simplified using the anticommutation rules:
\begin{equation}
    \bra{\psi_S}\hat{\gamma}_{-p\downarrow}\hat{\gamma}_{q\uparrow}\hat{\gamma}^{\dagger}_{\mathbf{k}_1\uparrow}\hat{\gamma}^{\dagger}_{-\mathbf{k}_2\downarrow}\ket{\psi_S} = 
    \delta_{\mathbf{k_1},\mathbf{q}}\delta_{\mathbf{k_2},\mathbf{p}}\, ,
\end{equation}
thus leading to:
\begin{equation}
    T^{(1)}_{qp} = \frac{i\sqrt{\gamma}m}{m_0 V} \int_{t_i}^{t}dt_1\,\widetilde{G}_{\mathbf{q}-\mathbf{p}}\widetilde{W}_{\mathbf{q}-\mathbf{p}}(t_1)M(q,p)e^{i\phi}e^{\frac{i}{\hbar}(E_{q}+E_{p})t_1}\, .
\end{equation}
This is the transition amplitude to a specific final state with fixed momenta $p$ and $q$. The transition probability from the BCS ground state to a state as in Eq.(\ref{pq}) is computed by taking the expectation value of the square modulus of the transition amplitude, according to Eq.(\ref{transprobboh}). In doing so, one has to compute the two-point correlator between the noise and its complex conjugate. Using the fact that $\widetilde{W}^*_{\mathbf{k}}(t)=\widetilde{W}_{-\mathbf{k}}(t)$ and Eq.(\ref{correlatortp}), one has that:
\begin{equation}\label{transprob}
  P_{qp} =\frac{\lambda m^2}{m_0^2 V}\widetilde{G}^2_{\mathbf{q}-\mathbf{p}}\,M^2(q,p)\,t \, ,
\end{equation}
where we assumed that $t_0=0$ and we performed integration over $t_1$. This probability grows linearly with time. 
By dividing Eq.(\ref{transprob}) by $t$, we obtain the transition rate, i.e. the rate at with which two quasiparticles with given momentum $p$ and $q$ are generated by the CSL noise. This transition rate will be the starting point to compute the generation rate in the next section, through which we can obtain the evolution of the quasiparticle occupation function $f(E)$, and thus the quasiparticle density. We present further calculations of the total generation rate of quasiparticles per unit volume in appendix \ref{thirdapp}.

\section{Quasiparticle density due the CSL noise}\label{phonon}
In the following, we consider a superconductor at thermal equilibrium and we neglect all sources of environmental noise. To account for thermal effects, we should consider the interaction of electrons with other electrons and with phonons, but since typically the electron-phonon interaction is dominant, we will consider only this one. 
The electron-phonon interaction comprises three main physical processes: quasiparticle scattering (both by emission and absorption of a phonon), quasiparticle recombination (by the emission of a phonon) and quasiparticle generation (by the absorption of a phonon).

The kinetic equation for the quasiparticle occupation function describes how $f(E)$ redistributes over time because of the above processes.
It contains the rates of the electron-phonon processes, and a generation rate given by an external source \cite{martinis2009energy,goldie2012non}:
\begin{equation}\label{diffeq}
\begin{split}
    &\frac{d f(E)}{dt}= \gamma_{g}^\text{\tiny ext}(E)+ \frac{\gamma_0}{\Delta^3}\int_{E}^{\infty}dE'S(E,E')\times\\
    &\,\,\,\,\,\,\,\,\,\,\,\,\,\,\,\,\,\,\,\,\,[(\bar{f}(E))f(E')(N(E'-E)+1)\\
    &\,\,\,\,\,\,\,\,\,\,\,\,\,\,\,\,\,\,\,\,-f(E)\bar{f}(E')N(E'-E)]\\
    &+\frac{\gamma_0}{\Delta^3}\int_{\Delta}^{E}dE' S(E,E')[(\bar{f}(E))f(E')N(E-E')\\
    &\,\,\,\,\,\,\,\,\,\,\,\,\,\,\,\,\,\,\,\,-f(E)\bar{f}(E')(N(E-E')+1)]\\
    &+\frac{\gamma_0}{\Delta^3}\int_{\Delta}^{\infty}dE'G(E,E')[(\bar{f}(E))\bar{f}(E')N(E+E')\\
    &\,\,\,\,\,\,\,\,\,\,\,\,\,\,\,\,\,\,\,\,-f(E)f(E')(N(E+E')+1)]\, ,
    \end{split}
\end{equation}
where $S(E,E')=(E-E')^2\rho(E')L^2(E,E')$, $G(E,E')=(E+E')^2\rho(E')M^2(E,E')$ and the rate $\gamma_0$ is a characteristic electron-phonon rate and it is a constant for a given material (for aluminum $1/\gamma_0=\tau_0 = 438$ns \cite{martinis2009energy}). The factor $N(\Omega)$ is the occupation function of phonons, which is taken to follow a Bose-Einstein distribution:
\begin{equation}\label{phononoccfunct}
    N(\Omega) =\frac{1}{e^{\Omega/k_B T_{ph}}-1}\, ,
\end{equation}
for a bath of phonons at temperature $T_{ph}$. Phonons are supposed to be in equilibrium at the refrigerator temperature and thus $N(\Omega)$ does not change in time.
The rate $\gamma_{g}^\text{\tiny ext}(E)$ is a generation rate per unit time due to external sources. The second and third terms in Eq.(\ref{diffeq}) are associated to thermal quasiparticle scattering, and the fourth term to thermal quasiparticle recombination and generation. 

One can check that when quasiparticles and phonons are in thermal equilibrium, i.e. when $f(E)$ is a Fermi-Dirac distribution and $N(\Omega)$ is a Bose-Einstein distribution with $T_{ph}=T$, the terms in every square brackets of Eq.(\ref{diffeq}) cancel. This is not surprising since, if $\gamma_{g}^\text{\tiny ext}=0$, i.e. there is no source of quasiparticles, the superconductor remains at equilibrium and the occupation function $f(E)$ does not change in time.

Similarly to the phonon case, the CSL Hamiltonian (\ref{cslhamgamintmain}) contains terms associated to scattering, recombination and generation of quasiparticles because of the interaction with the CSL noise. 
For temperatures below $100$mK, we checked that the generation rate is the only significant process. Let us then compute this rate when two quasiparticles, one at a fixed energy $E$ and the other at any energy $E'$, are created by the interaction with the CSL noise.

To do so, we start from the transition probability in Eq.(\ref{transprob}), integrating over one momentum:
\begin{equation}\label{startingappfourth}
    \gamma_g^\text{\tiny CSL}(E_{\mathbf{q}}) = \frac{\lambda (4\pi r_c)^{3/2} m^2}{m_0^2 (2\pi)^3}\int d^3p e^{-r_c^2(p-q)^2} M^2(p,q)\bar{f}(E_{\mathbf{p}})\, .
\end{equation}
Defining $x= E/\Delta$ and following the calculations in appendix \ref{fourthapp} we find:
\begin{equation}\label{cslgenratefin}
\begin{split}
        &\gamma_g^\text{\tiny CSL}(x) = \frac{m^2\lambda r_c}{2\sqrt{\pi}m_0^2}\frac{\sqrt{2m\Delta}}{\hbar} \frac{1}{\sqrt{s(x)+\beta}}e^{-\frac{\Delta}{k_B T_\text{\tiny CSL}}s(x)}\\
        &e^{-\frac{2T_F}{T_\text{\tiny CSL}}}\int_{1}^{\infty} dy \,e^{-\frac{\Delta}{k_B T_\text{\tiny CSL}}s(y)}\,e^{\frac{2\Delta}{k_B T_\text{\tiny CSL}}(\sqrt{(s(x)+\beta)(s(y)+\beta)}}\\
    &\times \rho(y)\frac{1}{2}\bigg(1-\frac{\sqrt{x^2-1}\sqrt{y^2-1}}{xy}+\frac{1}{xy}\bigg)\bar{f}(\Delta y)\, ,
    \end{split}
\end{equation}
where $s(x) =\sqrt{x^2-1}$, $\beta =\epsilon_F/\Delta$ and $k_B T_\text{\tiny CSL} = \hbar^2/(2m r_c^2)$.

\begin{figure}[H]
\begin{center}
\includegraphics[scale = 0.42]{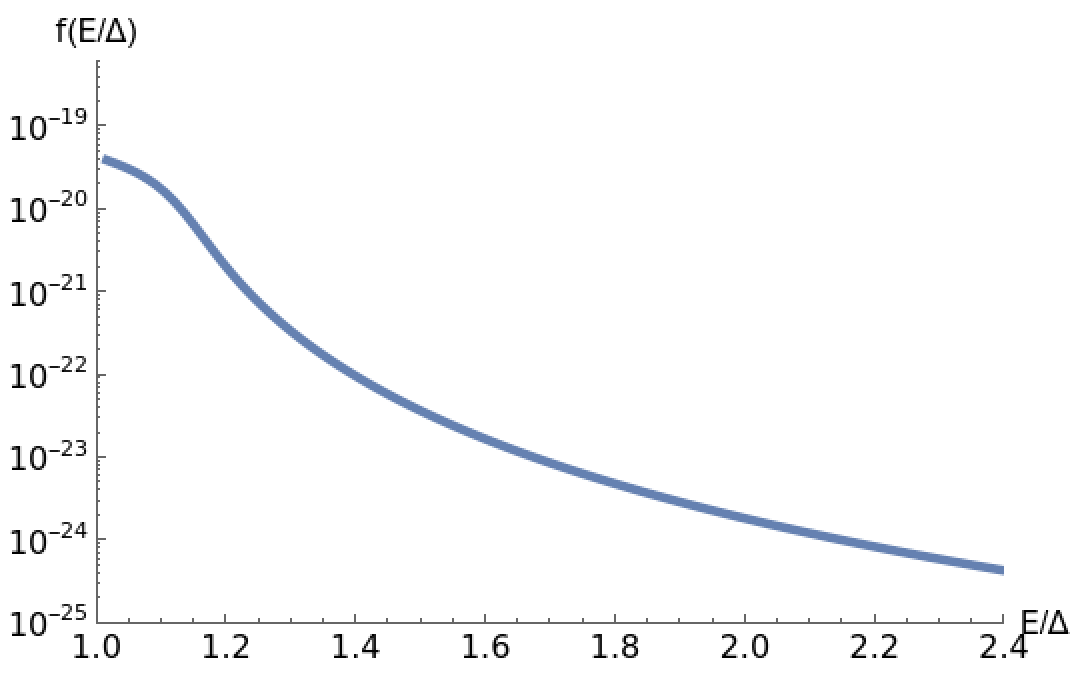}
\caption{Log plot of the quasiparticle occupation function obtained by solving Eq.(\ref{diffeq}) numerically for a starting equilibrium temperature of $20$ mK.
The evolved occupation function deviates significantly from the initial thermal distribution $f_{FD}$, here not shown because too small.
}\label{ssoccfunct}
\end{center}
\end{figure}

We solved numerically Eq.(\ref{diffeq}) with $\gamma_{g}^\text{\tiny ext} = \gamma_g^\text{\tiny CSL}$ and $\lambda = 10^{-10}$s$^{-1}$. 
Figure \ref{ssoccfunct} shows the evolved occupation function obtained numerically (blue solid line) for a starting equilibrium temperature of $20$ mK, which is a typical operational temperature of transmon qubits.
The steady state deviates strongly from the starting Fermi Dirac distribution, given by Eq.(\ref{fermidirac}) with $T= 20$ mK. The evolved occupation function can not be approximated by a thermal distribution at some effective temperature. More on the evolution of the quasiparticle occupation function can be found in appendix \ref{steadystate}.

The quasiparticle density $x^{\text \tiny{CSL}}_{qp}$ generated by CSL is obtained easily by using Eq.(\ref{qpdensity}) with $f(E)$ given by the evolved quasiparticle occupation function. $x^{\text \tiny{CSL}}_{qp}$ turns out to be $\sim 10^{-18}$. This quasiparticle density is $\sim 9$ orders of magnitude lower than the lowest reported experimental value of $x_{qp} = 10^{-9}$ \cite{serniak2019direct}.
This means that other environmental noise sources are yet dominant with the current techological implmentation of the transmon qubits. Nevertheless we can use the CSL quasiparticle density to get a limit on the coherence time of transmon qubits.

\section{Possibility of testing CSL with superconducting devices}\label{testingCSL}
Our analysis not only leads to possible limits on the performances of superconducting devices imposed by CSL, but also suggests that CSL dissipation could be tested with such devices.

We compare the electron-phonon generation and recombination rates with the CSL generation rate computed in section \ref{phonon} [see Eq.(\ref{cslgenratefin})], to find the temperatures for which the CSL generation rate dominates over the electron-phonon generation and/or recombination rates.

The recombination and generation rates enter  the last term of Eq.(\ref{diffeq}).
The electron-phonon recombination rate $\gamma_{r}^{e-ph}(E)$ is the inverse lifetime for a quasiparticle at some given energy $E$ to recombine with another quasiparticle of any energy $E'$ by emitting a phonon of energy $E+E'$. Its expression is given by\cite{chang1977kinetic,martinis2009energy}: 
\begin{equation}\label{recrateE}
\begin{split}
\gamma_{r}^{e-ph}(E)= \frac{\gamma_0}{\Delta^3}\int_{\Delta}^{\infty}dE'& (E+E')^2\rho(E')\bigg(1+\frac{\Delta^2}{EE'}\bigg)\\
&(N(E+E')+1)f(E')\, .
\end{split}
\end{equation}

We now briefly outline the different contributions appearing in the above integral. The term $(E+E')^2$ is the square energy of the emitted phonon during the recombination process. The normalized superconducting density of states $\rho(E')$ (see Eq.(\ref{scdensity})) appears since we are integrating over $E'$. The term $(1+\Delta^2/(EE'))$ is the coherence factor squared $M^2(E,E')$ in Eq.(\ref{coherencefact}).
The factor $N(\Omega)$ is the occupation function of phonons, a Bose-Einstein distribution Eq.(\ref{phononoccfunct}) at fixed $T= T_{ph}$.
The recombination rate is proportional to the quasiparticle density because of the factor $f(E')$ in the integrand \cite{martinis2009energy}. This implies that recombination mediated by phonons is slow when the density of quasiparticles is small. 

The electron-phonon generation rate $\gamma_{g}^{e-ph}(E)$ is the inverse lifetime for a quasiparticle at some given energy $E$ to be generated with another quasiparticle of any energy $E'$ by absorbing a phonon of energy $E+E'$ \cite{chang1977kinetic,martinis2009energy}. Its expression is given by:
\begin{equation}
\begin{split}
  \gamma_g^{e-ph}(E) =  \frac{\gamma_0}{\Delta^3}\int_{\Delta}^{\infty}dE'&(E+E')^2\rho(E')\bigg(1+\frac{\Delta^2}{EE'}\bigg)\\
  &\bar{f}(E')N(E+E')\, .
  \end{split}
\end{equation}
where $\bar{f}(E) = 1-f(E)$.

\begin{figure}[H]
\begin{center}
\includegraphics[scale = 0.40]{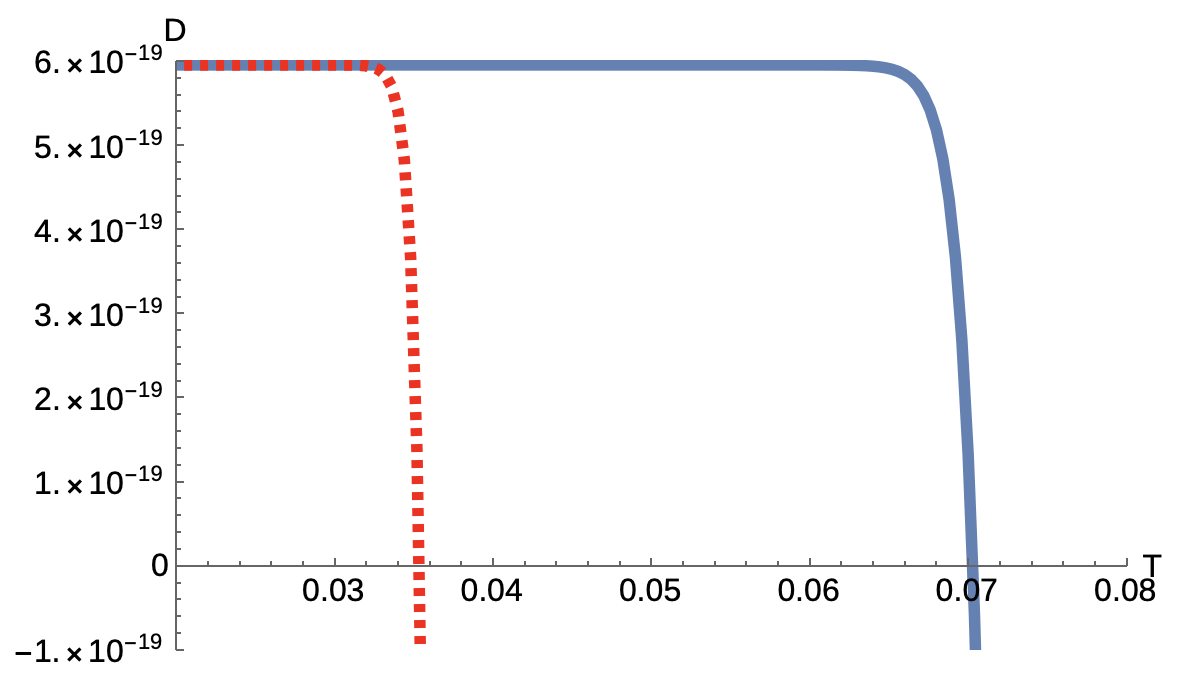}
\caption{Difference between the CSL generation rate and the electron-phonon generation rate $D_1$ (blue solid line), and difference between the CSL generation rate and the electron-phonon recombination rate $D_2$ (red dashed line), as functions of temperature. $D_1$ is positive for temperatures lower than $\sim 70$mK and $D_2$ is positive for temperatures lower than $\sim 35$mK. This implies that for temperatures lower than $\sim 35$mK the CSL generation rate is dominant over both the electron-phonon generation and recombination rates. This regime is accessible by experiments, since current refrigerators reach temperatures down to $\sim 20$ mK. Isolating a superconductor at these temperatures could lead to the detection of quasiparticles generated by the CSL noise.}\label{tempcomparison}
\end{center}
\end{figure}

In figure \ref{tempcomparison} we fix the energy at $\Delta$ and plot the two differences $D_1= \gamma_g^\text{\tiny CSL}(\Delta,T)-\gamma_g^{e-ph}(\Delta,T)$ and $D_2=\gamma_g^\text{\tiny CSL}(\Delta,T)-\gamma_r^{e-ph}(\Delta,T)$ as $T$ varies from $20$ mK to $80$mK. We assume that quasiparticles and phonons are in equilibrium so that $T_{ph}= T$ for every temperature $T$.

We see that the CSL generation rate is dominant over the electron-phonon generation rate ($D_1>0$) for temperatures lower than $\sim 70$mK, and similarly $D_2>0$ for temperatures lower than $\sim 35$mK. In particular, at temperatures lower than $35$mK the CSL generation rate is many orders of magnitude larger than the electron-phonon generation rate. Since a superconducting sample can be easily cooled down to $\sim 20$mK, this regime is already accessible by experiments. 
Of course, in order to perform a meaningful test of CSL, one should isolate the system from any other external source of quasiparticles. Note however that the CSL generation rate is itself very small $\sim 10^{-18}$ s$^{-1}$ (computed through Eq.(\ref{cslgenratefin}) at $E=\Delta$), thus making this kind of experiment challenging.

A straightforward way to perform such an experiment is by measuring the quasiparticle subgap current through a Josephson junction, which as we saw is a key component in superconducting qubits.
When two identical superconductors at absolute zero are linked together with an insulator, at voltages smaller than $V<2\Delta/e$ no quasiparticle current flows, since no quasiparticle state is excited. At finite temperature and $V<2\Delta/e$, there will 
be a finite current of quasiparticles, called quasiparticle subgap current \cite{milliken2004subgap}.
The formula for the subgap current at the voltage difference $V$ is given by:
\begin{equation}
\begin{split}
    I_{qp} &=\frac{1}{eR_N}\int dE\, \rho(E)\rho(E+eV)(f(E)-f(E+eV))\\
    & = \frac{I_c}{\Delta}\int dE\, \rho(E)\rho(E+eV)(f(E)-f(E+eV))\\
    & = I_c \int dx\, \rho(x)\rho(x+eV/\Delta)(f(x)-f(x+eV/\Delta))
\end{split}
\end{equation}
where $R_N$ and $I_c$ are, respectively, the normal state resistance and the critical current of the junction. Note that we used the Ambegaokar-Baratoff relation $R_N e = \pi\Delta/2 I_c\approx \Delta/I_c$ \cite{barone1982physics}, with $e$ the electron charge, in the second line, and in the third line we performed the substitution $x= E/\Delta$.

Plugging our computed occupation function in this equation, for a junction with critical current $I_c \sim 10^{-4}A$, we find a quasiparticle current of the order of $I^{\text \tiny{CSL}}_{qp}\sim 10^{-24} A$. This value has to be compared to the experimental values which are of the order of $I^{\text \tiny{exp}}_{qp}\sim 10^{-12} A$ \cite{milliken2004subgap}. A CSL quasiparticle current $I^{\text \tiny{CSL}}_{qp}$ of such intensity is extremely difficult to detect, as experiments that measure the quasiparticle current have a sensitivity of the order of pA.

The value of the CSL quasiparticle subgap current depends on the dimension of the Josephson junction and on the volumes of the two linked superconductors. Superconducting qubits have small superconducting volumes and thin Josephson junctions to work in the quantum regime and achieve a coherent control of the tunneling of single Cooper pairs, but larger volumes imply a larger number of quasiparticles.

Also, one can consider superconducting materials different from aluminum. The critical temperature $T_c$ of the superconductor might play a role for two reasons. First, a superconductor with lower $T_c$ may have a lower temperature at which the subgap current saturates because of external sources \cite{de2011number,milliken2004subgap}. Second, superconductors with higher $T_c$ increase the CSL quasiparticle generation as Eq.(\ref{ratevol}) depends on $\Delta =1.76k_B T_c$.

It is worthwhile noticing that the perturbative approach developed in this work can be easily extended to higher orders. Second order effects could be relevant for CSL induced tunneling events of quasiparticles through Josephson junctions. 

\section{Limits set by CSL in superconducting quantum computers}\label{fundmentallimits}
We use the quasiparticle density obtained in section \ref{phonon}, to estimate the ultimate limits set by CSL dissipation on the performances of quantum computers based on transmon qubits \cite{koch2007charge}.
First, we estimate the coherence time allowed by CSL for a single qubit. Next, we estimate how this would limit the performance of a quantum computer made of $N$ of these qubits.


State of the art transmon qubits have a relaxation time $T_1$ (the inverse of the relaxation rate $\Gamma_1$) of the order of $10-100$ $\mu$s \cite{kjaergaard2020superconducting}. 
A future goal is to achieve greater $T_1$ in order to have more reliable qubits, but this is experimentally challenging. Relaxation can be driven by many loss channels, such as radiation losses \cite{corcoles2011protecting}, dielectric losses \cite{wang2015surface}, two-level fluctuators in the junction materials \cite{klimov2018fluctuations} and by the excess of quasiparticles in the superconducting materials \cite{catelani2011relaxation,glazman2021bogoliubov,martinis2009energy}. 
The relaxation rate depends on every of these loss channel affecting the qubit \cite{vepsalainen2020impact}.

For a transmon qubit, the contribution to the relaxation rate due to quasiparticles depends linearly on the normalized quasiparticle density $x_{qp}$ \cite{vepsalainen2020impact}:
\begin{equation}\label{ratequbitqp}
    \Gamma_1 = \sqrt{\frac{2\omega_q\Delta}{\pi^2\hbar}}x_{qp}
\end{equation}
where $\omega_q$ is the frequency of the given qubit (in \cite{vepsalainen2020impact} $\omega_q = 2\pi\times3.48$ GHz).

Given $x^{\text \tiny{CSL}}_{qp}\sim 10^{-18}$ that we found in the previous section, we have $\Gamma^{\text \tiny{CSL}}_1 \approx 10^{-6}$s$^{-1}$. This is $\sim 10$ orders of magnitude larger than the CSL reduction rate estimated in section \ref{reduction}, showing that CSL dissipation is way more effective than the direct collapse process in corrupting superpositions of transmon qubits basis states. Note that $T^{\text \tiny{CSL}}_1$ is anyhow $10$ orders of magnitude larger than the $T_1 =100$ $\mu$s of current transmon qubits, implying that CSL dissipation would not influence effectively the performance of a single qubit. 

By knowing the limit on the coherence time of a single transmon qubit, we can estimate the limit on the performances of a quantum computer with $N$ of these transmon qubits. 
Decoherence of each qubit accumulates during the operational time of the quantum computer, eventually spoiling the quantum computation.

Before proceeding, we mention that schemes to recover from errors during a quantum computation, so called quantum error correction schemes \cite{gottesman2010introduction}, are planned to be implemented in future devices.
However scaling a quantum computer with an implemented error correction scheme is not an easy task.

In the near term, so called Noisy Intermediate Scale Quantum (NISQ) \cite{preskill2018quantum} devices could become a viable tool, whose qubits are subject to noise, without any implemented error correction scheme. 
We then consider the effect of CSL on the performance of a NISQ transmon quantum computer composed of $N$ qubits, each with a decay rate $\Gamma^{\text \tiny{CSL}}_1$.
Quantum algorithms on such a device might require the storage of a maximally entangled state of all these $N$ qubits \cite{mooney2019entanglement}, whose total decoherence rate $\Gamma_{tot}$ scales with $N$ and depends on the individual decay rate $\Gamma_1$ of the qubits \cite{palma1996quantum,ischi2005decoherence}. 

Since the volume of a typical transmon qubit is hundreds of microns cube \cite{martinis2009energy,wang2014measurement}, we roughly and conservatively estimate that the distance among such qubits is greater than $r_c$. Recalling Eq.(\ref{redcsladler}), one thus has that the CSL rate scales linearly with $N$: $\Gamma^{\text \tiny{CSL}}_{tot}=N \times \Gamma^{\text \tiny{CSL}}_1$.
Thus, the CSL limit on the decoherence time of a single qubit, $T^{\text \tiny{CSL}}_1 \approx 10^6$ s, gives a limit on $T^{\text \tiny{CSL}}_{tot} = 1/\Gamma^{\text \tiny{CSL}}_{tot}$, the decoherence time of a quantum computer of $N$ qubits. 
In order to gain a good fidelity of the output of the quantum computation, i.e. an output as close as possible to the desired result, the operational time of the quantum algorithm should be substantially smaller than the total decoherence time. The operational time of a quantum algorithm can be naively defined as $T_{op}= n_g \times t_g$, where $n_g$ is the number of quantum gates and $t_g$ is the time to implement one of these gate operations. State of the art devices have a $t_g \approx 10-100$ ns.
Then, what one requires is that:
\begin{equation}\label{limitesutop}
T_{op} \ll T^{\text \tiny{CSL}}_{tot} \Rightarrow n_g \ll \frac{T^{\text \tiny{CSL}}_1}{ t_g\,N}
\end{equation}
Note the resemblance of this condition to the so called "rule of thumb" metric for the performance of a quantum computer found in \cite{salm2020criterion}, if one defines $\epsilon =T^{\text \tiny{CSL}}_1/t_g$. 

As outlined in \cite{salm2020criterion}, it is not easy to estimate how smaller $T_{op}$ should be with respect to $T^{\text \tiny{CSL}}_{tot}$. Having a $T_{op}\approx T^{\text \tiny{CSL}}_{tot}$ implies that the algorithm is highly probable to fail \cite{Moll_2018}. 
Note moreover that the naive estimate we gave for $T_{op}$ is optimistic, as many factors can contribute to increase it. The most important examples are the restricted connectivity between qubits in a quantum chip and the possibility of implementing only the native gates allowed by the hardware \cite{salm2020criterion}. The first factor introduces additional swap gates so that two-qubit gates are performed only between physically connected qubits, the second factor requires the decomposition of the gates of the algorithm into those belonging to the native gate set, adding extra gate operations. 
Given these considerations, we will conservatively require that $T_{op}\approx 10^{-3}\times T^{\text \tiny{CSL}}_{tot}$. 

State-of-the-art transmon quantum computers with $N\approx 10^2$ would have a total decoherence time induced by CSL of $T^{\text \tiny{CSL}}_{tot}\sim 10^4$ s. Thus the total operational time to reach a sufficiently accurate result is $T_{op} \approx 10$s, and by having $t_g = 10^{-7}$ s, this allows a maximum of $n_g \approx 10^8$ operations. A $N = 10^3$ transmon quantum computer, targeted by IBM in 2023, allows for $n_g = 10^7$. Many important quantum algorithms such as molecular simulation, Shor algorithm for prime number factorization \cite{reiher2017elucidating,kutin2006shor}, which require respectively at least $N \sim 10^2$, $n_g \sim 10^{14}$, $N \sim 10^3$, $n_g \sim 10^9$ \cite{childs2018toward} will be corrupted by the CSL noise.
NISQ transmon quantum computers will scale up, eventually reaching the milestone of $N=10^6$, which is thought to be the number of qubits necessary to apply error correction schemes. Note that in this case the total time allowed by CSL without error correction will decrease to $T^{\text \tiny{CSL}}_{tot}\sim 10^{-3}$s.
\begin{figure}[H]
\begin{center}
\includegraphics[scale = 0.4]{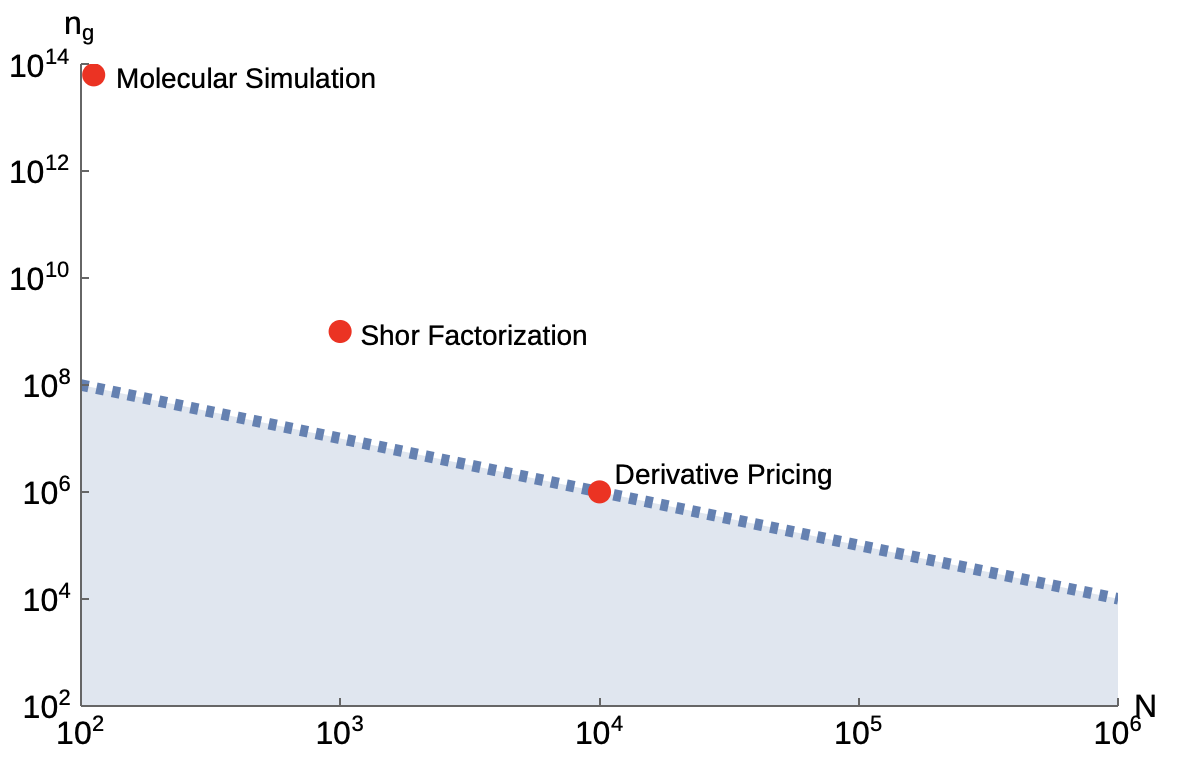}
\caption{Log-log plot of the number $n_g$ of allowed gates as a function of the number $N$ of qubits (blue dotted line) as dictated by Eq.(\ref{limitesutop}) with $T_1^{\text \tiny{CSL}}=10^{6}$s and fixed $t_g = 10^{-7}$s. The colored area under this curve corresponds to algorithms that use resources (i.e. number of qubits and number of gates) that allow to reach a good fidelity of the output. The three red dot correspond to the resources necessary to complete algorithms such as Shor factorization, molecular simulation and derivative pricing.}\label{limitscsl}
\end{center}
\end{figure}
\noindent This would allow for a maximum number of $n_g \approx 10^4$ gate operations, which is very far for what is needed to complete quantum algorithms for real life application \cite{shor1999polynomial,childs2018toward}. 

Figure \ref{limitscsl} summarizes these results: we plot the number of allowed gates as a function of the number of qubits (blue dotted line) as dictated by Eq.(\ref{limitesutop}). Quantum algorithms that exploit a number of qubits and a number of gates that stay under this curve (colored blue area) can reach a good fidelity of the output under the influence of CSL. The three red dots show the resources, estimated in \cite{childs2018toward} and \cite{chakrabarti2021threshold}, in order to complete important quantum algorithms such as prime number factorization, molecular simulation and derivative pricing \cite{chakrabarti2021threshold}. The points corresponding to Shor factorization and molecular simulation lie outside of the colored area: this means that CSL may spoil quantum computation with transmon NISQ quantum computers and also stresses the need of scaling up the devices with the possibility of performing quantum error correction or other error mitigation techniques. In fact even if the CSL limit on the coherence time of a single qubit is of the order of $10^6$s, which seems an extraordinary long time, the performances of a quantum computer could be spoiled without any scheme to recover from errors.

We conclude by pointing out that the values we obtained should be taken as rough estimates of the fundamental limitation imposed by collapse models on the performances of transmon quantum computers. To get a more accurate result, one needs to focus on a specific algorithm and calculate in detail the resources, in terms of number of qubits and of quantum gates, needed to complete it with a sufficiently good fidelity of the output. Indeed the number of quantum gates may vary a lot depending on different aspects \cite{salm2020criterion}, such as the specific physical hardware used and the algorithm to be solved.

\section{Conclusions and Outlook}\label{conclusion}
We showed how the CSL model affects superconducting quantum computers. The intrinsic localization of superpositions dictated by collapse models leaves the superposition of basis states of transmon qubits intact for very long times. However CSL contributes to decoherence also in an indirect way: dissipation induced by the CSL noise perturbs the superconducting material and leads to the generation of quasiparticles. These accumulate over time inside the volume of the device leading to relaxation at a rate proportional to their density. We estimated the quasiparticle density due to CSL by adding the CSL generation rate of quasiparticles to the kinetic equation for the quasiparticle occupation function.
We solved the kinetic equation numerically to find its steady state solution. With this calculation we obtained a lower quasiparticle density than the experimental one, so we conclude that other environmental noise sources are currently giving the dominant contribution to the experimental excess of quasiparticles. 

However, assuming one can eliminate environmental noises, the CSL excess quasiparticle density still limits the coherence time of 
a transmon quantum computer. CSL dissipation does not influence significantly a single qubit, as the coherence time allowed by CSL dissipation is of the order of $10^6$s, but it is relevant for a NISQ quantum computer composed of many qubits in which a complex quantum algorithm is run. 
Indeed the total decoherence time $T^{\text \tiny{CSL}}_{tot}$ of a quantum computer is inversely proportional to the number of qubits $N$ stacked together in its processor. This implies that, as the technology scales up to a larger $N$, less gate operations can be applied before the state of the quantum computer is corrupted by noise, as shown in figure \ref{limitscsl}. We showed that important algorithms such as prime number factorization and molecular simulation could be spoiled by CSL. Our analysis is performed by assuming that there is no quantum error correction scheme implemented in the devices. More accurate results could be obtained by focusing on a specific algorithm to find the resources needed to solve it, possibly including quantum error correction.

We further explored the possibility of testing CSL models with superconducting devices. The fact that the experimental values for the density of quasiparticle and for the subgap quasiparticle current are bigger than the CSL ones, implies that the detection of CSL effects is currently beyond the experimental sensitivity of superconducting devices, for which other environmental sources are dominant. 
We do not exclude that testing CSL models may be possible in the future as the technology develops \cite{cardani2021reducing}, given the importance that superconducting devices have for quantum computing. Our result show that when a superconducting sample is sufficiently shielded against environmental noises, CSL quasiparticles could be detected at the current refrigerators temperature.

\begin{acknowledgments}
Discussions with M. Paternostro and A. Varlamov are gratefully ackowledged. AB and LF acknowledge financial support from the H2020 FET Project TEQ (Grant No. 766900). AT and AB acknowledge support form the CNR/RS (London) project "Testing fundamental theories with ultracold atoms". AB acknowledges the Foundational Questions Institute and Fetzer Franklin Fund, a donor advised fund of Silicon Valley Community Foundation (Grant No. FQXi-RFP-CPW- 2002), INFN and the University of Trieste.
\end{acknowledgments}

\appendix

\section{BCS theory}\label{firstapp}
The BCS Hamiltonian of the system of electrons is given by:
\begin{equation}
    \hat{H}_{BCS} = \sum_{\mathbf{k}\sigma} \xi_{\mathbf{k}} \hat{c}_{\mathbf{k}\sigma}^{\dagger}\hat{c}_{\mathbf{k}\sigma} + \sum_{\mathbf{k}k'} U_{\mathbf{k}k'}\hat{c}_{\mathbf{k}\uparrow}^{\dagger}\hat{c}_{-\mathbf{k}\downarrow}^{\dagger}\hat{c}_{-\mathbf{k}'\downarrow}\hat{c}_{\mathbf{k}'\uparrow}
\end{equation}
where $\xi_{\mathbf{k}} = \hbar k^2/2m -\epsilon_F$ is the 
energy measured with respect to the Fermi energy $\epsilon_F$ and $U_{\mathbf{k}k'}$ is the interaction potential. The first term of the Hamiltonian is the kinetic energy while the second potential term couples pairs of different momenta $k$ and $k'$. The ground state of the Hamiltonian is the BCS ground state (\ref{gsbcs}) in the main text.
The form of the BCS Hamiltonian is involved since the potential term contains four fermionic operators. A simplified form is found through the mean field procedure. One defines
\begin{equation}
    a_{\mathbf{k}} = \langle \hat{c}_{\mathbf{k}\uparrow}^{\dagger}\hat{c}_{-\mathbf{k}\downarrow}^{\dagger} \rangle
\end{equation}
and assumes that the fluctuations $(\hat{c}_{\mathbf{k}\uparrow}^{\dagger}\hat{c}_{-\mathbf{k}\downarrow}^{\dagger}-a_{\mathbf{k}})$ are negligible. Then the following substitution:
\begin{equation}
     \hat{c}_{\mathbf{k}\uparrow}^{\dagger}\hat{c}_{-\mathbf{k}\downarrow}^{\dagger} = a_{\mathbf{k}} + \big(\hat{c}_{\mathbf{k}\uparrow}^{\dagger}\hat{c}_{-\mathbf{k}\downarrow}^{\dagger} -a_{\mathbf{k}}\big)
\end{equation}
(and its conjugate) is performed in the BCS Hamiltonian. By keeping terms up to first order in the fluctuations, one obtains the Bogoliubov Hamiltonian:
\begin{equation}
\begin{split}
     \hat{H}_{B} &= \sum_{\mathbf{k}\sigma} \xi_{\mathbf{k}} \hat{c}_{\mathbf{k}\sigma}^{\dagger}\hat{c}_{\mathbf{k}\sigma} \\
     &+ \sum_{\mathbf{k}k'} U_{\mathbf{k}k'}[a_{\mathbf{k}'}\hat{c}_{\mathbf{k}\uparrow}^{\dagger}\hat{c}_{-\mathbf{k}\downarrow}^{\dagger}+a_{\mathbf{k}}\hat{c}_{-\mathbf{k}'\downarrow}\hat{c}_{\mathbf{k}'\uparrow} -a_{\mathbf{k}} a_{\mathbf{k}'}]\, , 
\end{split}
\end{equation}
and defining $\Delta_{\mathbf{k}} = \sum_{\mathbf{k}'}U_{\mathbf{k}k'}a_{\mathbf{k}'}$ the Hamiltionian becomes:
\begin{equation}
       \hat{H}_{B} = \sum_{\mathbf{k}\sigma} \xi_{\mathbf{k}} \hat{c}_{\mathbf{k}\sigma}^{\dagger}\hat{c}_{\mathbf{k}\sigma} - \sum_{\mathbf{k}} \Delta_{\mathbf{k}}[\hat{c}_{\mathbf{k}\uparrow}^{\dagger}\hat{c}_{-\mathbf{k}\downarrow}^{\dagger}+\hat{c}_{-\mathbf{k}\downarrow}\hat{c}_{\mathbf{k}\uparrow}+ a_{\mathbf{k}}]\, .   
\end{equation}
At this point one performs the Bogoliubov transformation:
\begin{equation}
\begin{split}
    &\hat{\gamma}_{\mathbf{k}\uparrow} = u_{\mathbf{k}} \hat{c}_{\mathbf{k}\uparrow}-v_{\mathbf{k}} e^{i\phi} \hat{c}^{\dagger}_{-\mathbf{k}\downarrow} \,\,\,\,\,\,\hat{\gamma}_{-\mathbf{k}\downarrow} = v_{\mathbf{k}} e^{i\phi} \hat{c}^{\dagger}_{\mathbf{k}\uparrow}+u_{\mathbf{k}} \hat{c}_{-\mathbf{k}\downarrow}\\
     &\hat{\gamma}^{\dagger}_{\mathbf{k}\uparrow} = u_{\mathbf{k}} \hat{c}^{\dagger}_{\mathbf{k}\uparrow}-v_{\mathbf{k}} e^{-i\phi} \hat{c}_{-\mathbf{k}\downarrow} \,\,\,\,\,\, \hat{\gamma}^{\dagger}_{-\mathbf{k}\downarrow} = v_{\mathbf{k}} e^{-i\phi} \hat{c}_{\mathbf{k}\uparrow}+u_{\mathbf{k}} \hat{c}^{\dagger}_{-\mathbf{k}\downarrow}\, ,
\end{split}
\end{equation}
and the inverse Bogoliubov transformation is given by: 
\begin{equation}\label{boginverse}
\begin{split}
    &\hat{c}_{\mathbf{k}\uparrow} = u_{\mathbf{k}} \hat{\gamma}_{\mathbf{k}\uparrow}+v_{\mathbf{k}} e^{i\phi}\hat{\gamma}^{\dagger}_{-\mathbf{k}\downarrow} \,\,\,\,\,\, \hat{c}_{-\mathbf{k}\downarrow} = u_{\mathbf{k}} \hat{\gamma}_{-\mathbf{k}\downarrow}-v_{\mathbf{k}} e^{i\phi}\hat{\gamma}^{\dagger}_{\mathbf{k}\uparrow} \\
     &\hat{c}^{\dagger}_{\mathbf{k}\uparrow} = u_{\mathbf{k}} \hat{\gamma}^{\dagger}_{\mathbf{k}\uparrow}+v_{\mathbf{k}} e^{-i\phi}\hat{\gamma}_{-\mathbf{k}\downarrow}  \,\,\,\,\,\,  \hat{c}^{\dagger}_{-\mathbf{k}\downarrow} = u_{\mathbf{k}} \hat{\gamma}^{\dagger}_{-\mathbf{k}\downarrow}-v_{\mathbf{k}} e^{-i\phi}\hat{\gamma}_{\mathbf{k}\uparrow}\, .
\end{split}
\end{equation}
Substituting Eqs.(\ref{boginverse}) in the Bogoliubov Hamiltonian one finds:
\begin{equation}
\begin{split}
    \hat{H}_B & = \sum_{\mathbf{k}} \big[ \xi_{\mathbf{k}}(u_{\mathbf{k}}^2-v_{\mathbf{k}}^2) + 2\Delta_{\mathbf{k}} u_{\mathbf{k}} v_{\mathbf{k}} \big]\big[\hat{\gamma}^{\dagger}_{\mathbf{k}\uparrow}\hat{\gamma}_{\mathbf{k}\uparrow} + \hat{\gamma}^{\dagger}_{-\mathbf{k}\downarrow}\hat{\gamma}_{-\mathbf{k}\downarrow} \big]\\
    & + \sum_{\mathbf{k}} \big[ 2\xi_{\mathbf{k}} u_{\mathbf{k}} v_{\mathbf{k}} - \Delta_{\mathbf{k}} (u_{\mathbf{k}}^2 - v_{\mathbf{k}}^2) \big]\big[\hat{\gamma}^{\dagger}_{\mathbf{k}\uparrow}\hat{\gamma}^{\dagger}_{-\mathbf{k}\downarrow} + \hat{\gamma}_{-\mathbf{k}\downarrow}\hat{\gamma}_{\mathbf{k}\uparrow}\big]\\
    & + \sum_{\mathbf{k}} \big[ 2\xi_{\mathbf{k}} v_{\mathbf{k}}^2 - 2\Delta_{\mathbf{k}} u_{\mathbf{k}} v_{\mathbf{k}} + \Delta_{\mathbf{k}} a_{\mathbf{k}}\big] \, .
\end{split}
\end{equation}
This expression contains undesired terms of the type $\hat{\gamma}\hat{\gamma}$ and $\hat{\gamma}^{\dagger}\hat{\gamma}^{\dagger}$ so the coefficients of these terms are set to zero \cite{grosso2013solid, tinkham2004introduction}:
\begin{equation}
   2\xi_{\mathbf{k}} u_{\mathbf{k}} v_{\mathbf{k}} - \Delta_{\mathbf{k}} (u_{\mathbf{k}}^2 - v_{\mathbf{k}}^2) = 0 \, .
\end{equation}
This condition together with the normalization condition $u_{\mathbf{k}}^2+v_{\mathbf{k}}^2 = 1$ gives:
\begin{align}
    & u_{\mathbf{k}}^2 =\frac{1}{2}\bigg(1+\frac{\xi_{\mathbf{k}}}{E_{\mathbf{k}}}\bigg) &v_{\mathbf{k}}^2=\frac{1}{2}\bigg(1-\frac{\xi_{\mathbf{k}}}{E_{\mathbf{k}}}\bigg) 
\end{align}
The quasiparticle energies are given by:
\begin{equation}
 \xi_{\mathbf{k}}(u_{\mathbf{k}}^2-v_{\mathbf{k}}^2) + 2\Delta_{\mathbf{k}} u_{\mathbf{k}} v_{\mathbf{k}} = \sqrt{\xi_{\mathbf{k}}^2 + \Delta_{\mathbf{k}}^2}\, ,
\end{equation}
and the Bogoliubov Hamiltionian can be expressed in the form:
\begin{equation}\label{bogham}
    \hat{H}_B = \sum_{\mathbf{k}} E_{\mathbf{k}} \big[\hat{\gamma}^{\dagger}_{\mathbf{k}\uparrow}\hat{\gamma}_{\mathbf{k}\uparrow} + \hat{\gamma}^{\dagger}_{-\mathbf{k}\downarrow}\hat{\gamma}_{-\mathbf{k}\downarrow} \big] + W_S \, ,
\end{equation}
with $E_{\mathbf{k}} = \sqrt{\xi_{\mathbf{k}}^2+\Delta_{\mathbf{k}}^2}$, and $W_S = \sum_{\mathbf{k}} \big[ 2\xi_{\mathbf{k}} v_{\mathbf{k}}^2 - 2\Delta_{\mathbf{k}} u_{\mathbf{k}} v_{\mathbf{k}} + \Delta_{\mathbf{k}} a_{\mathbf{k}}\big]$.
The superconducting gap parameters are given by:
\begin{equation}
\begin{split}
    \Delta_{\mathbf{k}} =& \sum_{\mathbf{k}'} U_{\mathbf{k}k'}a_{\mathbf{k}'} = \sum_{\mathbf{k}'} U_{\mathbf{k}k'} \langle \hat{c}_{\mathbf{k}'\uparrow}^{\dagger}\hat{c}_{-\mathbf{k}'\downarrow}^{\dagger} \rangle\\
    &=  \sum_{\mathbf{k}'} U_{\mathbf{k}k'}u_{\mathbf{k}'}v_{\mathbf{k}'}\langle 1-\hat{\gamma}^{\dagger}_{\mathbf{k}'\uparrow}\hat{\gamma}_{\mathbf{k}'\uparrow} - \hat{\gamma}^{\dagger}_{-\mathbf{k}'\downarrow}\hat{\gamma}_{-\mathbf{k}'\downarrow}\rangle\\
    & = \sum_{\mathbf{k}'} U_{\mathbf{k}k'}u_{\mathbf{k}'}v_{\mathbf{k}'}(1-f(E_{\mathbf{k}'}))
    \end{split}
\end{equation}
and by substituting the expression of $u_{\mathbf{k}}$ and $v_{\mathbf{k}}$ one finds the usual self consistent equations:
\begin{equation}
    \Delta_{\mathbf{k}} = -\frac{1}{2}\sum_{\mathbf{k}'}U_{\mathbf{k}k'}\frac{\Delta_{\mathbf{k}'}}{E_{\mathbf{k}'}}(1-f(E_{\mathbf{k}'}))
\end{equation}
The BCS assumption is that $U_{\mathbf{k}k'}= -V$, a negative constant, for $k$ such that $|\xi_{\mathbf{k}}|<\hbar\omega_D$, and $U_{\mathbf{k}k'}=0$ otherwise. In this way one has:
\begin{equation}
\Delta_{\mathbf{k}} = \begin{cases}
\Delta  \,\, for \,\,|\xi_{\mathbf{k}}|<\hbar\omega_D \\
0 \,\,for\,\, |\xi_{\mathbf{k}}|>\hbar\omega_D
\end{cases}
\end{equation}
The equation for the gap becomes then:
\begin{equation}
\Delta = \frac{V}{2}\sum_{\mathbf{k}'}\frac{\Delta}{E_{\mathbf{k}'}}(1-2f(E_{\mathbf{k}'}))
\end{equation}
Simplifying the common $\Delta$ factor we are left with:
\begin{equation}\label{gapsmallT}
\begin{split}
   &1= \frac{V}{2}\sum_{\mathbf{k}'}\frac{1}{E_{\mathbf{k}'}}(1-2f(E_{\mathbf{k}'}))\\
   &1 = g(\epsilon_F)V\bigg(\int_{\Delta}^{\hbar\omega_D} \frac{dE}{\sqrt{E^2-\Delta^2}}-\int_{\Delta}^{\hbar\omega_D} dE \rho(E)\frac{1}{E}2f(E)\bigg)\\
  & 1\approx g(\epsilon_F)V\bigg(\ln{\frac{2\hbar\omega_D}{\Delta}}-x_{qp}\bigg)
\end{split}
\end{equation}
where in the third line we switched to an integration over the quasiparticle energies $E$.
The latter equation can be solved for the gap:
\begin{equation}
    \Delta = 2\hbar\omega_D e^{-1/g(\epsilon_F)V-x_{qp}}=\Delta(0) e^{-x_{qp}}\approx \Delta(0)(1-x_{qp})
\end{equation}
where $\Delta_0 = 2\hbar\omega_D e^{-1/g(\epsilon_F)}= 1.76k_B T_c$ is the superconducting gap with no quasiparticles, so at absolute zero.
Notice that the gap depends on the normalized quasiparticle density and therefore on the occupation function of quasiparticles. However for small occupation function, and for small enough temperatures, it is reasonable to approximate $\Delta = \Delta(0)$.

\section{CSL Model}\label{secondapp}

The CSL model is usually formulated in position space. The collapse of the wave function is described by a non-linear and stochastic interaction with a classical noise through the Itô equation:
\begin{equation}\label{csleqtotapp}
\begin{split}
    d\ket{\psi} =\bigg[&-\frac{i}{\hbar}\hat{H}dt+\frac{\sqrt{\lambda}}{m_0}\int d^3x \big(\hat{M}(\mathbf{x})- \langle \hat{M}(\mathbf{x})\rangle \big)dW_t(\mathbf{x})\\
   &-\frac{\lambda}{2m_0^2}\int d^3x\,d^3y \big(\hat{M}(\mathbf{x})- \langle \hat{M}(\mathbf{x})\rangle  \big)G(\mathbf{x}-\mathbf{y})\times\\
   &\,\,\,\,\,\,\,\,\,\,\,\,\,\,\,\,\,\,\,\,\,\,\,\big(\hat{M}(\mathbf{y})- \langle \hat{M}(\mathbf{y})\rangle  \big)dt\bigg]\ket{\psi}
\end{split}
\end{equation}
where $m_0$ is the nucleon mass, $\langle \,\cdot\, \rangle$ denotes the expectation value on the state $\ket{\psi}$ and $\hat{M}(\mathbf{x})$ is the mass density operator defined by:
\begin{equation}\label{moperator}
\hat{M}(\mathbf{x}) = \sum_j m_j \hat{a}_j^{\dagger}(\mathbf{x})\hat{a}_j(\mathbf{x})\, .
\end{equation}
The operators $\hat{a}_j^{\dagger}(\mathbf{x})$ and $\hat{a}_j(\mathbf{x})$ are the creation and annihilation operators at position $\mathbf{x}$ of a particle of type $j$ with mass $m_j$ (in our case we will have a single $m$ given by the mass of electrons). The $G(\mathbf{x}-\mathbf{y})$ in (\ref{csleqtotapp}) are Gaussian functions of the form:
\begin{equation}
    G(\mathbf{x}-\mathbf{y}) = \frac{1}{(4\pi r_c^2)^{3/2}}e^{-\frac{1}{4r_c^2}(x-y)^2}\, ,
\end{equation}
that characterize the statistical properties of the noise $W_t(\mathbf{x})$. Indeed, by calling $\xi_t(\mathbf{x}) = dW_t(\mathbf{x})/dt$, one has that $\mathbb{E}[\xi_t(\mathbf{x})]=0$, and the two point correlator $\mathbb{E}[\xi_t(\mathbf{x}),\xi_s(\mathbf{y})]=G(\mathbf{x}-\mathbf{y})\delta(t-s)$ where $\mathbb{E}[\cdot]$ denotes the stochastic average.
These properties together with the mass proportionality of $\hat{M}$ guarantee respectively localization in space and the amplification mechanism: the collapse rate of a body of $N$ constituents gets amplified linearly in $N$.
It is generally difficult to work directly with Eq. (\ref{csleqtotapp}), mainly because of its non-linearity. Since we are interested in expectation values we can use the simplified linear, but still stochastic, dynamic given by Eq.(\ref{evcslsimpl}) in the main text.
The CSL term in position space and in the Stratonovich form is given by:
\begin{equation}\label{CSLhamapp}
    \hat{H}_\text{\tiny CSL}= -\frac{\hbar\sqrt{\lambda}}{m_0} \int d^3x\, \xi_t(\mathbf{x})\hat{M}(\mathbf{x})\, .
\end{equation}
This term is related to the second term in equation (\ref{csleqtotapp}), but now it is linear because it does not contain $\langle\hat{M}(\mathbf{x})\rangle$ anymore. Moreover the third term of Eq.(\ref{csleqtotapp}) is not present.
These simplifications are possible because of the equivalence of Eqs.(\ref{csleqtotapp}) and (\ref{CSLhamapp}) at the statistical level: non-linearity effects are washed away when expectation values are computed.
The term (\ref{CSLhamapp}) is Fourier transformed to obtain Eq.(\ref{cslhamc}) of the main text: we work in the normalization volume $V$ to avoid any divergences and the position representation of the field operators is related to the momentum operators via:
\begin{align}
\hat{a}(\mathbf{x},s) &= \frac{1}{\sqrt{V}}\sum_{\mathbf{k}} e^{i\mathbf{k}\mathbf{x}}\hat{c}_{\mathbf{k}s}\\
\hat{a}^{\dagger}(\mathbf{x},s) &= \frac{1}{\sqrt{V}}\sum_{\mathbf{k}} e^{-i\mathbf{k}\mathbf{x}}\hat{c}^{\dagger}_{\mathbf{k}s}\, .
\end{align}

\section{Calculation of the total generation rate of quasiparticles per unit time and unit volume}\label{thirdapp}
We can have a first estimate of the CSL effects by computing the total rate $\Gamma$ of generation of quasiparticles per unit time and unit volume. This corresponds to performing a sum over momenta $p$ and $q$ on the transition probability (\ref{transprob}):
\begin{equation}
\begin{split}
      &\Gamma = \frac{\lambda m^2}{m_0^2 V}\sum_{q,p}\widetilde{G}^2_{\mathbf{q}-\mathbf{p}}\,M^2(q,p)\\
      &= \frac{\lambda m^2 V r_c}{4m_0^2\pi^{5/2}}\int dq q \int dp p\,\big[e^{-r_c^2(p-q)^2}-e^{-r_c^2(p+q)^2}\big]M^2(q,p)\,,
      \end{split}
\end{equation}
where in the second line we expressed the summations as integrals in spherical coordinates.

To make the integral adimensional, we perform the following substitutions: $\hbar p/\sqrt{2m\Delta} = x$ and $\hbar q/\sqrt{2m\Delta} = y$.
We recall that $\Delta_{\mathbf{k}} =\Delta$ for $|\xi_{\mathbf{k}}|<\hbar\omega_D$ and zero otherwise, which gives the following constraints for the modulus of $x$:
\begin{equation}
  A_- < x < A_+
\end{equation}
where $A_{\pm}=\sqrt{\frac{\epsilon_F \pm \hbar\omega_D}{\Delta}}$. The same applies to the modulus of $y$. Performing these substitutions we find:
\begin{equation}
\begin{split}
   \Gamma =& \frac{\lambda m^2 V r_c}{4m_0^2\pi^{5/2}}  \bigg(\frac{2m\Delta}{\hbar^2}\bigg)^2 \int_{A_-}^{A_+} dy \int_{A_-}^{A_+} dx \,xy\\
   &\big(e^{-\frac{2m\Delta r_c^2}{\hbar^2}(x-y)^2}-e^{-\frac{2m\Delta r_c^2}{\hbar^2}(x+y)^2}\big)\times\\
   &\bigg(1-\frac{(x^2-\beta)(y^2-\beta)}{[((x^2-\beta)^2+1)((y^2-\beta)^2+1)]^{\frac{1}{2}}}\\
   &+\frac{1}{[((x^2-\beta)^2+1)((y^2-\beta)^2+1)]^{\frac{1}{2}}}\bigg)\\
\end{split}
\end{equation}
where $\epsilon_F/\Delta = \beta$.
We can compare the different values of the adimensional parameters appearing in the above integral. We have that $\beta\sim 10^4$, $\frac{2m\Delta r_c^2}{\hbar^2}\sim 10^2$, $A_{-}^2 = \beta -\frac{\hbar\omega_D}{\Delta} \sim 10^4-10$ and $A_{+}^2 = \beta +\frac{\hbar\omega_D}{\Delta} \sim 10^4+10$. Then the Gaussian functions can be considered as Dirac deltas, and since we are integrating in two intervals for $x$ and $y$ where $x$ has the same sign of $y$, the second Gaussian gives no contribution. By exploiting the Dirac delta representation:
\begin{equation}
    \delta(t) = \lim_{\epsilon \to 0} \frac{1}{\epsilon\sqrt{2\pi}} e^{-\frac{1}{2\epsilon^2}t^2}\, ,
\end{equation} 
where in our case $\frac{1}{2\epsilon^2} = \frac{2m\Delta r_c^2}{\hbar^2}$, we are then left with:
\begin{equation}\label{ratevol}
\begin{split}
   &\Gamma = \frac{\lambda m^2 V}{4\pi^2 m_0^2}\bigg(\frac{\sqrt{2m}}{\hbar}\bigg)^3 \Delta^{2}\bigg(\sqrt{\epsilon_F} \int_{0}^{+\hbar\omega_D} d\xi \,\frac{1}{(\xi^2+\Delta^2)}\bigg)\\
   &= \frac{\lambda m^2 V}{8 m_0^2\pi}\bigg(\frac{\sqrt{2m}}{\hbar}\bigg)^3 \sqrt{\epsilon_F}\Delta\, ,
\end{split}
\end{equation}
where in the first line we already performed the change of variables from $x$ to $\xi$, and in the second line we approximated the integral to $\pi/2$ since it is equal to $\tan^{-1}(\hbar\omega_D/\Delta)$ with $\hbar\omega_D/\Delta >> 1$. 
Assuming that each quasiparticle is generated with energy $\Delta$, the total power density supplied by the CSL to a superconducting sample of volume $V$ is Eq.(\ref{ratevol}) multiplied by $\Delta$.
Plugging in these equations the parameters for aluminum ($\epsilon_F=11.6$eV and $\Delta =3.4\times 10^{-4}$eV), and a value $\lambda = 10^{-10}$s$^{-1}$, the total CSL generation rate is $\Gamma \approx 3\times 10^{-11}$s$^{-1}\mu$m$^{-3}$ and the power density $P_{tot} = 1\times 10^{-33}$W$\mu$m$^{-3}$. Other works \cite{martinis2009energy} estimated the total generation rate per unit volume and the power density that would account for the experimental quasiparticle density measured in transmon qubits. The values that we obtained are $\sim 14-16$ orders of magnitude smaller than the values estimated in \cite{martinis2009energy}, $\Gamma_g^{ext}= 2.4\times 10^3$s$^{-1}$ $\mu$m$^{-3}$ and $P_{tot}= 6\times 10^{-14}$ W$\mu$m$^{-3}$, thus showing that we can not attribute the current experimental excess of quasiparticles to the CSL noise, which is due to other sources. 
However we can neglect them, and compute the steady state quasiparticle density due to CSL, as we do in the main text. In order to do so, the information that the total generation rate gives is incomplete. Thermal processes and the CSL generation of quasiparticles contribute to the evolution of the occupation function $f(E)$ of quasiparticles. In this regime $f(E)$ redistributes over time to a steady state different from a thermal state, that enters Eq.(\ref{qpdensity}) to give the quasiparticle density due to CSL.

\section{Calculation of generation rate of quasiparticles per unit time}\label{fourthapp}
We start from Eq.(\ref{startingappfourth}) of the main text:
\begin{equation}
\begin{split}
    \gamma_g^\text{\tiny CSL}&(E_{\mathbf{q}}) = \frac{\lambda (4\pi r_c)^{3/2} m^2}{m_0^2 (2\pi)^3}\int d^3p e^{-r_c^2(p-q)^2} M^2(p,q)\bar{f}(E_{\mathbf{p}})\\
    & = \frac{\lambda m^2 (4\pi)^{3/2}r_c}{m_0^2 2(2\pi)^2}\int dp\frac{p}{q}\bigg(e^{-r_c^2(p-q)^2}-e^{-r_c^2(p+q)^2}\bigg)\\
    &\,\,\,\,\,\,\,\,\,\,\,\,\,\,\,\,\,\,\,\,\,\,\,\,\,\,\,\,\,\,\,\,\,\,\,\,M^2(p,q)\bar{f}(E_{\mathbf{p}})\, ,
    \end{split}
\end{equation}
where in the second line we expressed the integral in polar coordinates. We make the following substitutions: $q=\sqrt{2m}/\hbar(\sqrt{E^2-\Delta^2}+\epsilon_F)^{1/2}$ and $p=\sqrt{2m}/\hbar(\sqrt{E'^2-\Delta^2}+\epsilon_F)^{1/2}$ to obtain:
\begin{equation}
\begin{split}
        &\gamma_{g}^\text{\tiny CSL}(E) =\frac{m^2\lambda r_c}{2\sqrt{\pi}m_0^2}\frac{\sqrt{2m}}{\hbar} \frac{1}{\sqrt{\sqrt{E-\Delta^2}+\epsilon_F}}\\
        &\int_{\Delta}^{\infty} dE'
    \bigg(e^{-\frac{2mr_c^2}{\hbar^2}(\sqrt{\sqrt{E-\Delta^2}+\epsilon_F)}-\sqrt{\sqrt{E'-\Delta^2}+\epsilon_F)})^2}\\
    &-e^{-\frac{2mr_c^2}{\hbar^2}(\sqrt{\sqrt{E-\Delta^2}+\epsilon_F)}+\sqrt{\sqrt{E'-\Delta^2}+\epsilon_F)})^2}\bigg)\\
    &\times \rho(E')M^2(E,E')\bar{f}(E')\, .
    \end{split}
\end{equation}
The above expression is simplified by expanding the squares in the exponential and collecting the common factors:
\begin{equation}
\begin{split}
       \gamma_g^\text{\tiny CSL}(E) =& \frac{m^2\lambda r_c}{2\sqrt{\pi}m_0^2}\frac{\sqrt{2m}}{\hbar} \frac{1}{\sqrt{\sqrt{E^2-\Delta^2}+\epsilon_F}}e^{-\frac{2mr_c^2}{\hbar^2}\sqrt{E^2-\Delta^2}} \\
        &e^{-\frac{4mr_c^2\epsilon_F}{\hbar^2}}\int_{\Delta}^{\infty} dE' e^{-\frac{2mr_c^2}{\hbar^2}\sqrt{E^2_1-\Delta^2}}\\
        &
    \bigg(e^{\frac{2mr_c^2}{\hbar^2}2(\sqrt{(\sqrt{E^2-\Delta^2}+\epsilon_F)(\sqrt{E'^2-\Delta^2}+\epsilon_F)}}\\
    &-e^{-\frac{2mr_c^2}{\hbar^2}2(\sqrt{(\sqrt{E^2-\Delta^2}+\epsilon_F)(\sqrt{E'^2-\Delta^2}+\epsilon_F)}}\bigg)\\
    &\times \rho(E')M^2(E,E')\bar{f}(E')\, .
   \end{split}
\end{equation}

By defining $T_\text{\tiny CSL}$ through $k_B T_\text{\tiny CSL} = \hbar^2/(2m r_c^2)$, and making the substitution $x = E/\Delta$ and $y = E'/\Delta$, we get:
\begin{equation}
\begin{split}
        &\gamma_g^\text{\tiny CSL}(x) = \frac{m^2\lambda r_c}{2\sqrt{\pi}m_0^2}\frac{\sqrt{2m\Delta}}{\hbar} \frac{1}{\sqrt{s(x)+\beta}}e^{-\frac{\Delta}{k_B T_\text{\tiny CSL}}s(x)}\\
        &e^{-\frac{2T_F}{T_\text{\tiny CSL}}}\int_{1}^{\infty} dy \,e^{-\frac{\Delta}{k_B T_\text{\tiny CSL}}s(y)}\rho(y)M^2(x,y)\bar{f}(\Delta y))\\
    &\bigg(e^{\frac{2\Delta}{k_B T_\text{\tiny CSL}}(\sqrt{(s(x)+\beta)(s(y)+\beta)}}
    -e^{-\frac{2\Delta}{k_BT_\text{\tiny CSL}}(\sqrt{(s(x)+\beta)(s(y)+\beta})}\bigg)\, ,
    \end{split}
\end{equation}
where we called $s(x) = \sqrt{x^2-1}$.
We can further simplify this expression by neglecting the negative term in the last line because it is exponentially suppressed.
We can finally write:
\begin{equation}
\begin{split}
        &\gamma_g^\text{\tiny CSL}(x) = \frac{m^2\lambda r_c}{2\sqrt{\pi}m_0^2}\frac{\sqrt{2m\Delta}}{\hbar} \frac{1}{\sqrt{s(x)+\beta}}e^{-\frac{\Delta}{k_B T_\text{\tiny CSL}}s(x)}\\
        &e^{-\frac{2T_F}{T_\text{\tiny CSL}}}\int_{1}^{\infty} dy \,e^{-\frac{\Delta}{k_B T_\text{\tiny CSL}}s(y)}\,e^{\frac{2\Delta}{k_B T_\text{\tiny CSL}}(\sqrt{(s(x)+\beta)(s(y)+\beta)}}\\
    &\times \rho(y)\frac{1}{2}\bigg(1-\frac{\sqrt{x^2-1}\sqrt{y^2-1}}{xy}+\frac{1}{xy}\bigg)\bar{f}(\Delta y)\, .
\end{split}
\end{equation}
This is Eq.(\ref{cslgenratefin}) of the main text.

\section{Steady state solution}\label{steadystate}
In the main text, we have seen that the steady state solution to Eq.(\ref{diffeq}) when $\gamma_{g}^\text{\tiny ext}= 0$, is the Fermi Dirac distribution function.
This observation, together with the fact that the CSL injection rate $\gamma_{g}^\text{\tiny CSL}(E)$ is small, suggests the following procedure to approximate the steady state solution when $\gamma_{g}^\text{\tiny ext} = \gamma_{g}^\text{\tiny CSL}$: since the steady state solution without injection rate is a Fermi-Dirac distribution $f_{FD}(E,T)$, at some temperature $T$, when we add the CSL injection term, the modified steady state solution can be written as $f_{SS}(E) = f_{FD}(E,T)+\delta f(E)$ with $\delta f(E)$ a small perturbation.

We start from Eq.(\ref{diffeq}) and we neglect the rate terms associated to recombination and generation, since they are exponentially small. We plug $f_{SS}= f_{FD}+\delta f$ into Eq.(\ref{diffeq}), and we perform the substitutions $x=E/\Delta$, $y=E'/\Delta$ to get:
\begin{equation}
\begin{split}
    &\gamma_{g}^\text{\tiny CSL}(x)+ \gamma_0\int_{x}^{4}dy S(x,y)\big[(1-f_{FD}(x)-\delta f(x))\\
    &(f_{FD}(y)+\delta f(y))(N(y-x)+1)-(f_{FD}(x)+\delta f(x))\\
    &(1-f_{FD}(y))-(f_{FD}(x)+\delta f(x))(-\delta f(y))(N(y-x))\big] \\
    &+\gamma_0\int_{1}^{x}dy S(x,y)\big[(1-f_{FD}(x)-\delta f(x))(f_{FD}(y)+\delta f(y))\\
    &(N(x-y))-(f_{FD}(x)+\delta f(x))(1-f_{FD}(y)\\
    &-\delta f(y))(N(x-y)+1)\big] = 0\,.
    \end{split}
\end{equation}
The integration limit is $x= 4$ since we are interested in the low energy behaviour of the occupation function.

Developing the product of all these factors, the terms coming from the Fermi-Dirac distribution cancel out. Then, keeping terms up to first order in $\delta f$ one has:
\begin{equation}
\begin{split}
    &\gamma_{g}^\text{\tiny CSL}(x)+ \gamma_0\int_{x}^{4}dy S(x,y)\big[\bar{f}_{FD}(x)\delta f(y)(N(y-x)+1)\\
    &-\delta f(x)f_{FD}(N(y-x)+1)-\delta f(x)\bar{f}_{FD}(y)N(y-x)\\
    &+\delta f(x)f_{FD}(y)N(y-x)\big] +\gamma_0\int_{1}^{x}dy S(x,y)\big[\bar{f}_{FD}(x)\\
    &\delta f(y)N(x-y)-f_{FD}(y)\delta f(x)(N(x-y))-\delta f(x)\bar{f}_{FD}(y)\\
    &(N(x-y)+1)+f_{FD}(x)\delta f(y)(N(x-y)+1)\big] = 0\,.
    \end{split}
\end{equation}
where $\bar{f}_{FD}(x)=1-f_{FD}(x)$.
This is further simplified by neglecting terms proportional to $N(\Omega)$ (see Eq.(\ref{phononoccfunct}) of the main text), since they are exponentially small for the temperatures that we consider. In this way one ends up with:
\begin{equation}\label{sssol}
\begin{split}
    &\gamma_{g}^\text{\tiny CSL}(x)+ \gamma_0\int_{x}^{4}dy S(x,y)[(\bar{f}_{FD}(x))\delta f(y)(N(y-x)+1)\\
    &\,\,\,\,\,\,\,\,\,\,\,\,\,\,\,\,\,\,\,\,-\delta f(x)f_{FD}(y)(N(y-x)+1)]\\
    &+\gamma_0\int_{1}^{x}dy S(x,y)[(f_{FD}(x))\delta f(y)(N(x-y)+1)\\
    &\,\,\,\,\,\,\,\,\,\,\,\,\,\,\,\,\,\,\,\,-\delta f(x)\bar{f}_{FD}(y)(N(x-y)+1)]= 0\, 
    \end{split}
\end{equation}
where $\bar{f}_{FD}=(1-f_{FD})$.

From now on, we will approximate $\bar{f}_{FD}(x) = 1-f_{FD}(x)\approx 1$ and $N(\Omega)+1 \approx 1$, because at milliKelvin temperatures and in the considered energy interval $[\Delta,4\Delta]$, $f_{FD}(E)\sim e^{-E/k_bT}\approx 0$ and  $N(\Omega)\sim e^{-\Omega/k_bT}\approx 0$. The above equation thus simplifies to:
\begin{equation}\label{hofinitoidee}
\begin{split}
    &\gamma_{g}^\text{\tiny CSL}(x)+ \gamma_0\int_{x}^{4}dy S(x,y)[\delta f(y)-\delta f(x)f_{FD}(y)]\\
    &+\gamma_0\int_{1}^{x}dy S(x,y)[(f_{FD}(x))\delta f(y)-\delta f(x)] = 0\, .
    \end{split}
\end{equation}
To further simplify this expression, we note that, for milliKelvin temperatures, we can neglect the addends of the form $\delta f(x)f_{FD}(y)$ and $f_{FD}(x)\delta f(y)$, since we computed the corresponding integrals and they turned out to be small quantities with respect to the other terms in Eq.(\ref{hofinitoidee}). Finally one has that:
\begin{equation}
\begin{split}
    &\gamma_{g}^\text{\tiny CSL}(x)+ \gamma_0\int_{x}^{4}dy S(x,y)\delta f(y)\\
    &-\gamma_0\int_{1}^{x}dy S(x,y)\delta f(x) = 0\, ,
    \end{split}
\end{equation}
which can be inverted to find the following equation for $\delta f(x)$:
\begin{equation}\label{sssolapprox}
\begin{split}
\delta f(x) &= \frac{\gamma_{g}^\text{\tiny CSL}(x)}{\gamma_0\int_{1}^{x}dy S(x,y)}+ \frac{\int_{x}^{4}dy S(x,y)\delta f(y)}{\gamma_0\int_{1}^{x}dy S(x,y)}\\
&\approx  \frac{\gamma_{g}^\text{\tiny CSL}(x)}{\gamma_0\int_{1}^{x}dy S(x,y)}\, .
\end{split}
\end{equation}
This shows that, by knowing the CSL generation rate, we can compute the correction to the Fermi-Dirac distribution, and thus the steady state solution of Eq.(\ref{diffeq}) without the need to solve it.
The approximation in the second line of Eq.(\ref{sssolapprox}) is justified because if we substitute $\delta f(y) = \frac{\gamma_{g}^\text{\tiny CSL}(y)}{\gamma_0\int_{1}^{y}dz S(y,z)}$, in the second term of Eq.(\ref{sssolapprox}) we obtain a small correction to the first term. 
In the next section we check the validity of this approximation scheme.

\section{Validity of the approximation scheme}
\begin{widetext}

\begin{figure}[H]
\begin{center}
\includegraphics[scale = 0.31]{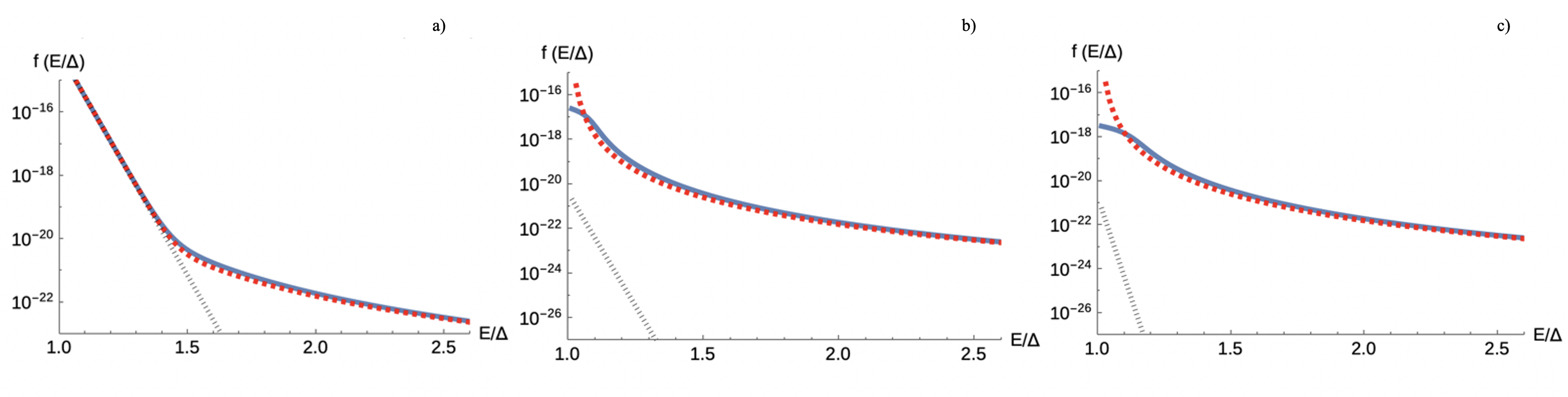}
\caption{Log plot of the quasiparticle occupation function obtained solving Eq.(\ref{diffeq}) numerically (blue solid line), of the the analytical expression $f_{SS}=f_{FD}+\delta f$ with $\delta f$ given by Eq.(\ref{sssolapprox}) (red dashed line) and of the starting Fermi-Dirac distribution $f_{FD}$ (black dotted line) for a starting equilibrium temperature of $65$ mK (a), $45$ mK (b) and $25$ mK (c). 
In a) $\delta f$ is a small perturbation to $f_{FD}$ only for energies lower than $\sim 1.5\Delta$, where the red dashed line representing $f_{SS}$ is close to the black dotted line representing $f_{FD}$. Despite this, the red dashed line is very close to the blue solid line, meaning that $f_{SS}$ is a good approximation to the occupation function obtained by numerical simulation.
In b) the correction $\delta f \gg f_{FD}$, since the red dashed line representing $f_{SS}$ is many orders of magnitude larger than the black dotted line representing $f_{FD}$. The same applies for $T=25$mK in c).
Analogously to a), for both b) and c) the red dashed line representing the analytical solution $f_{SS}$ is a good approximation to the blue solid line representing the occupation function obtained by numerical simulation. Note that in c) we multiplied by $10^{16}$ the initial $f_{FD}$ to have the same interval in the vertical axis of b). 
}\label{figurafinale1}
\end{center}
\end{figure}

\end{widetext}

To test the validity of this approximation scheme, we have chosen three representative starting equilibrium temperatures ($65$ mK, $45$ mK and $25$ mK) to check two aspects. 
First, we compared $f_{SS} = f_{FD}+\delta f$ with $f_{FD}$, to check if $\delta f$ is indeed a small perturbation to $f_{FD}$. 
Second, we compared $f_{SS}$ with the numerical solution of (\ref{diffeq}) (with $\gamma_{g}^\text{\tiny ext}(E)=\gamma_g^\text{\tiny CSL}(E)$), to check if $f_{SS}$ is a good approximation to the steady state solution of Eq.(\ref{diffeq}). We summarize the results in figure \ref{figurafinale1}, where we plot the occupation function obtained with the numerical simulation (blue solid line), the analytical expression $f_{SS}=f_{FD}+\delta f$ (red dashed line) and the initial Fermi Dirac distribution $f_{FD}$ (black dotted line), for a starting equilibrium temperature of $65$ mK (a), $45$ mK (b) and $25$ mK (c). 

In figure \ref{figurafinale1}a the perturbation $\delta f$ is small with respect to $f_{FD}$ only for energies lower than $\sim 1.5\Delta$, as the red dashed line representing $f_{SS} = f_{FD}+\delta f$ is close to the black dotted line representing $f_{FD}$ only in this energy interval.
Nevertheless the analytical expression $f_{SS}$ (red dashed line) is close to the numerical result represented by the blue solid line in the whole energy interval.
This means that, despite the fact that we are not allowed to treat $\delta f$ as a small perturbation to $f_{FD}$, $f_{SS}$ is a good approximation to the steady state solution of Eq.(\ref{diffeq}).

In figure \ref{figurafinale1}b the analytical solution (red dashed line) is many orders of magnitude larger than the initial Fermi-Dirac occupation function (black dotted line) in the whole energy interval. Thus, in this case $\delta f$ is not a small perturbation to $f_{FD}$. The same applies to figure \ref{figurafinale1}c.
Note however that, analogously to figure \ref{figurafinale1}a, in both figures \ref{figurafinale1}b and \ref{figurafinale1}c the numerical result (blue solid line) is well approximated by $f_{SS}$ (red dashed line). 

In conclusion, for the temperatures considered, even when $\delta f$ in Eq.(\ref{sssolapprox}) is not a small perturbation to $f_{FD}$, we can use $f_{SS} = f_{FD} + \delta f$ as the steady state solution of Eq.(\ref{diffeq}).

Note moreover that, the fact that for $T=45$ mK and $T=25$ mK, the correction $\delta f \gg f_{FD}$ implies that $f_{SS} = f_{FD}+ \delta f \approx \delta f$. That is to say, $\delta f$ given by Eq.(\ref{sssolapprox}), is itself the good approximation to the steady state solution of Eq.(\ref{diffeq}).
As $\delta f$ turns out to be almost independent of temperature for milliKelvin temperatures, this suggests that the steady state solution for any $T<45$ mK, can always be well approximated by $\delta f$. This fact was confirmed by additional simulations (here not shown) with starting equilibrium temperatures at $T=10$ mK and $T=5$ mK, for which the numerical result was still well approximated by $f_{SS}\approx \delta f$.

\bibliography{Paper.bib}

\end{document}